\documentclass[useAMS]{mn2e}

\usepackage{times}
\usepackage{amssymb}
\usepackage{rotating}
\usepackage{graphicx,epstopdf}
\usepackage{longtable}
\usepackage{array}
\usepackage{multirow}
\usepackage{tabularx}
\usepackage{caption}
\usepackage{amsmath} 
\usepackage{color}
\usepackage{siunitx}
\usepackage{float}
\usepackage[dvipsnames]{xcolor}
\usepackage{tabularx}
\usepackage{colortbl}
\usepackage{gensymb}
\usepackage{physics}
\newcommand{\pf}{\textsc{propfluc}}
\newcommand{\ro}{$r_{\text{o}}$}
\newcommand{\ri}{$r_{\text{i}}$}
\newcommand{\rd}{$r_{\text{d}}$}
\newcommand{\rbw}{$r_{\text{bw}}$}
\newcommand{\rg}{R$_{\text{g}}$}

\newcommand{\thickhline}{%
    \noalign {\ifnum 0=`}\fi \hrule height 1pt
    \futurelet \reserved@a \@xhline
}

\title[XTE J1550-564: challenges to propagating fluctuations]{Cross-spectral modelling of the black hole X-ray binary XTE J1550-564: challenges to the propagating fluctuations paradigm}
\author[S. Rapisarda, A. Ingram and M. van der Klis]{S. Rapisarda$^1$, A. Ingram$^1$ and M. van der Klis$^1$
\\
$^{1}$Anton Pannekoek Institute for Astronomy, University of Amsterdam, Science Park 904, 1098XH Amsterdam, Netherlands\\
}
\date{Accepted for publication in MNRAS}
\pagerange{\pageref{firstpage}--\pageref{lastpage}} \pubyear{2015}

\begin{document}


\pagerange{\pageref{firstpage}--\pageref{lastpage}} \pubyear{2017}

\maketitle
\topmargin = -0.5cm
\label{firstpage}

\begin{abstract}
Timing properties of black hole X-ray binaries in outburst can be modeled with mass accretion rate fluctuations propagating towards the black hole. Such models predict time lags between energy bands due to propagation delays. First application of a propagating fluctuations model to black hole power spectra showed good agreement with the data. Indeed, hard lags observed from these systems appear to be in agreement with this generic prediction. Our \pf{} code allows to simultaneously predict power spectra, time lags, and coherence of the variability as a function of energy. This was successfully applied to \textit{Swift} data on the black hole MAXI J1659-152, fitting jointly the power spectra in two energy bands and the cross-spectrum between these two bands. In the current work, we attempt to to model two high signal to noise \textit{Rossi X-ray Timing Explorer} (RXTE) observations of the black hole XTE J1550-564. We find that neither observation can be adequately explained by the model even when considering, additionally to previous \pf{} versions, different propagation speeds of the fluctuations. After extensive exploration of model extensions, we tentatively conclude that the quantitative and qualitative discrepancy between model predictions and data is generic to the propagating fluctuations paradigm. This result encourages further investigation of the fundamental hypotheses of the propagating fluctuations model. We discuss some of these hypotheses with an eye to future works.
\end{abstract}


\section{Introduction}
\label{sec:int}
Low mass black hole X-ray binaries (BHBs) are optimal laboratories for studying the physics of accretion. To investigate the geometry of, and the mechanism regulating, accretion, it is necessary to observe the spectral and timing characteristics of the source. During the rising phase of an outburst, these characteristics radically change on time scale of days/weeks (e.g. Remillard \& McClintock 2006; Belloni 2010; Gilfanov 2010). At the beginning of the outburst, when the luminosity is still low, the energy spectrum of the source is dominated by a non-thermal component usually modeled with a hard power law ($\Gamma \approx 1.5$). In this phase (also known as low-hard state, LHS), the power spectrum of the source is characterized by broad band noise (rms $\approx 30 \%$) and by a strong quasi periodic oscillation (QPO, e.g. Casella, Belloni \& Stella 2005 and references therein). As the source luminosity increases, the power law softens ($\Gamma \approx 2.5$) and the energy spectrum shows a significant thermal component that can be modeled by a multitemperature blackbody peaking at $\approx$ 1 keV. This component becomes dominant at maximum luminosity (high-soft state, HSS). In this transition, all the characteristic frequencies of the power spectral features increase (e.g. Wijnands \& van der Klis 1998; Homan et al. 2001) and the amplitude of both broad band noise and QPO decreases. The transition from LHS to HSS does not occur in the same way for all the BHBs, for example, in some cases, the very high state (VHS) is observed, which is characterized by high luminosity, a significant blackbody component, and a soft power law dominating the emission (e.g. Miyamoto et al. 1991; McClintock \& Remillard 2006). \\
Assuming that each spectral component corresponds to a certain physical region, the changes described above indicate important variations in the geometry of the accreting system. If each region produces variability depending on its physical characteristics, the correlation between the varying emission in different energy bands is an important tool to study both the different regions and their mutual interaction. In particular, it is known that BHBs show a positive (hard) lag between variations in soft and hard energy band and that the amplitude of this lag depends both on Fourier frequency and energy.\\  
The truncated disc model (e.g. Esin, McClintock \& Narayan 1997; Done, Gierli{\'n}ski \& Kubota 2007) is a good candidate for explaining the spectral transition described above: an optically thick geometrically thin disc is truncated at a certain radius \ro{} (truncation radius) larger than the last stable circular orbit (ISCO) and, inside this radius, accretion takes place through an optically thin (optical depth $\tau \sim$ 1.5) geometrically thick hot flow. Disc photons illuminating the hot flow are Compton up-scattered by the hot flow high energy electrons. This interaction both cools down the hot flow and produces the observed hard power law emission (Thorne \& Price 1975; Sunyaev \& Truemper 1979). At the beginning of the outburst, the disc is truncated far away from the the BH, so that the X-ray spectrum is dominated by the non-thermal emission coming from the hot flow. As the mass accretion rate increases, the truncation radius moves towards the BH, the luminosity of disc photons illuminating the hot flow increases cooling it down (the power law softens) and the dominant spectral component becomes the thermal multitemperature blackbody disc emission.  \\
Looking at the short time scale variability (luminosity variations with time scale from ms to $\approx$ 100s), it is possible to explain the timing properties of the source considering mass accretion rate fluctuations stirred up in every ring of the accreting region and propagating towards the BH (Lyubarskii 1997). Propagating mass accretion rate fluctuations have been claimed to naturally explain the linear rms-flux relation observed in BHBs (Uttley \& McHardy 2001; Uttley et al. 2005), the fact that the variability is observed over several orders of magnitude in frequency (Lyubarskii 1997), and the hard lag between soft and hard emission (Kotov et al. 2001; Arevalo \& Uttley 2006), but quantitative tests of these claims have so far been incomplete. \\
The reason for this is that, even though the truncated disc model and the mass accretion rate fluctuations scenario may explain some of the observed phenomenology, a physical model that can self-consistently reproduce the energy spectrum, the power spectrum, and the correlation between different energy bands, has not yet been constructed. The model \pf{} (Ingram \& Done 2011, 2012, hereafter ID11, ID12; Ingram \& van der Klis 2013, hereafter IK13; Rapisarda et al. 2016, hereafter RIKK16) is intended as a first step in this direction: the model is based on the truncated disc geometry and assumes that mass accretion rate fluctuations, stirred up both in the hot flow and in the disc, propagate towards the BH on a local viscous time scale. In combination with this process of propagating fluctuations, the entire hot flow precesses because of frame dragging around the BH (Lense-Thirring precession) producing the QPO (Stella \& Vietri 1999; Fragile et al. 2007; Ingram et al. 2016). The model outputs are power spectra in two different energy bands and the cross-spectrum between these bands. This implies that \pf{} can predict the phase lag and the correlation between the two bands.\\
RIKK16 applied the model to the BH MAXI J1659-152 using \textit{Swift} data. In their study, \pf{} predictions showed to be in reasonable agreement with the data, but the data showed only small phase lags associated with the variability. In order to test \pf{} predictions and, ultimately, the validity of the propagating fluctuations hypothesis, we studied high s/n XTE J1550-564 observations showing significant phase lags.\\  
XTE J1550-564 is a BH first detected with the All-Sky monitor (ASM; Levine et al. 1996) on board of the Rossi X-ray Timing Explorer (RXTE) on September 7th 1998 (Smith 1998). Since its discovery, the source has gone into outburst 4 times, but the first outburst (1998-1999) remains the most luminous and best monitored outburst of the source ever observed with RXTE. The rising part of this outburst (first 12 days from the discovery) is characterized by a transition from LHS state to VHS (Sobczak et al. 2000; Kubota \& Done 2004). The power spectrum of the source in this phase shows broad band noise and a strong QPO signal (Cui et al. 1999, Remillard et al. 2002). In our study we analyzed 12 observations from the beginning of the outburst, in particular we performed spectral analysis in a full energy band (1.94-20.30 keV) and timing analysis in two energy bands (1.94-12.99 keV and 13.36-20.30 keV, soft and hard band respectively). For testing \pf{}, we selected two observations (two and seven days after the discovery, respectively) characterized by different count rate and power spectral shape in the soft and hard band. We jointly fitted soft and hard band power spectra, and the cross-spectrum between these bands. \\
In Sec. \ref{sec:up} we present a number of updates to \pf{} intended to deal with the more complex variability characteristics of XTE J1550-564. In particular we add the option of extra variability in the inner part of the hot flow, and two new parameters describing the damping strength and the propagation velocity of the fluctuations. Sec. \ref{sec:obs} and \ref{sec:modfit} describe the techniques used for reducing the data and the \pf{} fit results, respectively. For both observations, the \pf{} prediction disagrees with the data for any acceptable combination of fit parameters. In Sec. \ref{sec:dis} we discuss the implication of this result and possible, more complex, physical scenarios not considered in \pf{}. In Sec. \ref{sec:con} we summarize our results.

\section{PROPFLUC updates}
\label{sec:up}
\pf{} (ID11, ID12, IK13) assumes that mass accretion rate fluctuations propagate through a truncated disc/hot flow geometry, where the entire hot flow is precessing because of frame dragging in the vicinity of a rotating BH (Lense-Thirring precession). RIKK16 improved the model in order to consider variability generated in and propagating from the truncated disc through the hot flow, and to simultaneously fit power spectra in two different energy bands and the complex cross-spectrum between these bands. Here, we introduce three extra updates to the model: 1) we consider the hypothesis of extra variability stirred up in the inner part of the hot flow; 2) we introduce a new parameter for taking into account the damping of mass accretion rate fluctuations as they propagate towards the BH; 3) we introduce a multiplicative factor regulating the propagation speed of the fluctuations. These updates are intended to make the \pf{} prescription resemble more closely a real accretion flow and extend the model capabilities in order to fit the complex variability observed in XTE J1550-564 (and in general in BHBs).  

\subsection{Summary of the \pf{} model}
\begin{figure} 
\center
\includegraphics[scale=0.4,angle=0]{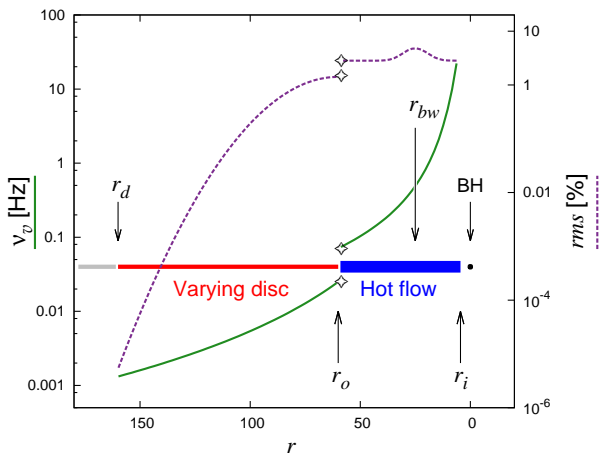}
\caption{Viscous frequency (green solid line) and amount of variability produced by each ring (purple dashed line) versus the radial coordinate covering truncated disc and hot flow. The variability profile is Gaussian around the transition radius \rbw{}. This extra variability in the hot flow, coupled with the variability stirred up in and propagating from the disc, produces a three hump power spectrum.}
\label{fig:nurms}
\end{figure}
\label{sec:extra}
The accreting region assumed by \pf{} consists of an optically thick geometrically thin disc and a optically thin geometrically thick hot flow (red and blue horizontal bars in Fig. \ref{fig:nurms}, respectively). The region in the disc contributing to the variability extends from the truncation radius \ro{} up to a radius \rd{} smaller than the outer edge of the disc. The hot flow extends from an inner radius \ri{} out to \ro{}. The inner radius \ri{} is equal to or larger than the innermost stable circular orbit (ISCO), so that $r_\text{d} > r_\text{o} > r_\text{i} > r_{ISCO}$. We adopt the convention that lowercase $r$ corresponds to radial coordinate scaled by gravitational radius: $r = R/{R_{g}}$, where $R_g = GM/c^2$. \\
For numerical purposes, both disc and hot flow are split into rings whose boundaries are equally logarithmically spaced. The model resolution, i.e. the number of rings per radial decade $N_{dec}$, is the same for both disc and hot flow. Each ring produces a Lorentzian spectrum of mass accretion rate fluctuations with characteristic frequency equal to the local viscous frequency. These fluctuations propagate towards the BH on a time scale equal to the local viscous time scale. The radial viscous frequency profile has thus the double role of determining the characteristic frequency of fluctuations stirred up in each ring and the propagation speed of these fluctuations.\\
In the disc, the viscous frequency depends on radius as in a Shakura-Sunyaev disc with constant viscosity parameter and scale-height: $\nu_{v, disc} (r)= \nu_{d, max} ( r / r_o )^{-3/2}$ (Shakura \& Sunyaev 1973). In this expression, $\nu_{d, max}$ is the maximum viscous frequency in the disc, obtained at \ro{}, and is one of the model parameters (see RIKK16, Appendix A, for details). \\
In the hot flow, the viscous frequency is described by a smoothly broken power law, i.e. two different power law indices set the viscous frequency profile for rings close and far from the BH respectively, with a smooth transition at the so-called \textit{bending wave} radius (ID12, IK13, Rapisarda et al. 2014). Because at each radius the viscous frequency is inversely proportional to the surface density (Frank, King \& Raine 2002), this radius marks a transition between two different density regimes in the hot flow: the surface density increases smoothly with radius at large radii, while at small radii, the surface density drops off towards the BH because of viscous torques. Numerical simulations (Fragile et al. 2007) show that a similar transition occurs in the wavelength of the bending waves (which propagate warps throughout the hot flow and so influence its shape), suggesting that the surface density transition radius is set by the shape of the bending waves (Ingram et al. 2009). This transition/bending wave radius, $r_{bw}$, is a model parameter and regulates how the material is distributed in the hot flow. \\
The green solid curve of Fig. \ref{fig:nurms} shows the viscous frequency in both the disc and the hot flow. The star symbols on the curve highlight the discontinuity of the viscous frequency trend at the truncation radius. It is this discontinuity of the physical properties of the propagating region that produces a double hump power spectrum (RIKK16). The power spectrum resulting from propagating mass accretion rate fluctuations depends also on the amount of variability produced by each ring. This is described by the model parameter $F_{var}$ = $\sigma(r) \sqrt{N_{dec}}$ (the fractional variability per radial decade), where $\sigma(r)$ is the amount of variability fractional rms injected at each radius. In the disc, $\sigma(r)$ is a Gaussian peaking at the truncation radius \ro{} (the amplitude and the width of the Gaussian are both model parameters), while in the hot flow it is assumed to be constant (ID12, Rapisarda et al. 2014, and RIKK16). 

\subsection{Extra variability in the flow}
\begin{figure} 
\center
\includegraphics[scale=0.45,angle=270]{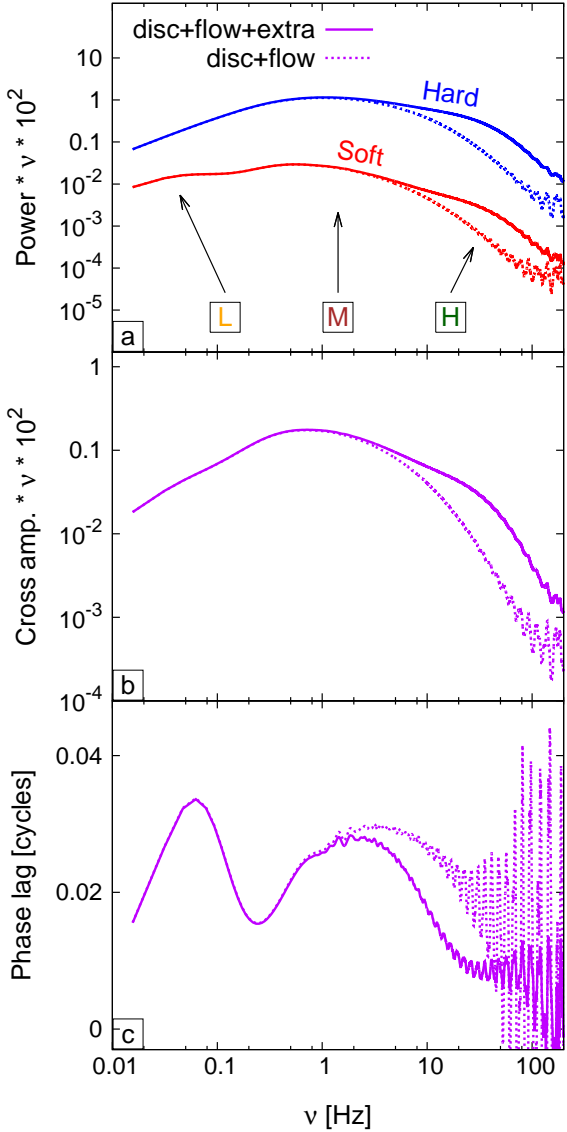}
\caption{Soft (red line) and Hard (blue line) power spectrum (a), cross-spectral amplitude (b), and phase lag (c) computed considering mass accretion rate fluctuations propagating in the hot flow + disc (dashed line) and adding extra variability in the hot flow (solid line). In the latter case $\textsc{propfluc}$ produces a three-hump power spectrum combining disc variability (L, low frequency hump), hot flow variability (M, main hump), and extra variability in the hot flow (H, high frequency hump).}
\label{fig:humps}
\end{figure}

Numerical simulations show that extra high-frequency continuum variability may be produced in the inner part of a tilted accretion flow (Fragile \& Blaes 2008; Henisey, Blaes \& Fragile 2012). The mechanism generating this extra variability is related to the changes in the accreting regime of the flow around the bending wave radius. Here we introduce a prescription to account for this extra variability in \pf{}: $\sigma(r)$ in the hot flow is now a Gaussian peaking at the transition radius \rbw{}. The amplitude and the width of this Gaussian, $N_{extra}$ and $\Delta r$ respectively, are both model parameters.\\
The purple dashed line of Fig. \ref{fig:nurms} shows $\sigma$ for both the disc and the hot flow. Fig. \ref{fig:humps} shows the result of including extra variability in the inner part of the hot flow. The dotted line is the two-hump power spectrum resulting from mass accretion rate fluctuations produced both in the disc and in the hot flow and propagating towards the BH. Including extra variability in the hot flow, we obtain a third, high-frequency hump in the power spectrum (label H in Fig. \ref{fig:humps}a), so a three-hump power spectrum. Because the inner part of the hot flow emits mostly in the hard band, we do not observe large phase lags associated with this third hump. However, the presence of the high frequency hump partly suppresses the lag associated with the hump produced by mass accretion rate fluctuations propagating only in the hot flow (main hump, M), as extra variability dilutes its cross-spectrum (see Fig. \ref{fig:humps}c around 10 Hz). The fringes observed at high frequency (see Fig. \ref{fig:humps} above $\approx$ 10 Hz) are the result of interference between contributions from different rings.\\
With the additional hypothesis of extra variability in the hot flow, it is now possible to produce a double hump power spectrum in two different ways: combining variability from the disc and the hot flow (labels L and M in Fig. \ref{fig:humps}a) and variability coming only from the hot flow (labels M and H in Fig. \ref{fig:humps}a). However, while by adjusting the \pf{} parameters it is possible to reproduce a double hump shape in the power spectrum in two different ways, the two configurations (L-M and M-H) produce a different phase lag profile depending on the spectral properties of the source, i.e. on how thermal and non-thermal emission is distributed between the selected hard and soft bands.
  
\subsection{The damping factor}
\begin{figure} 
\center
\includegraphics[scale=0.4,angle=0,trim=0cm 0 0.5cm 0,clip]{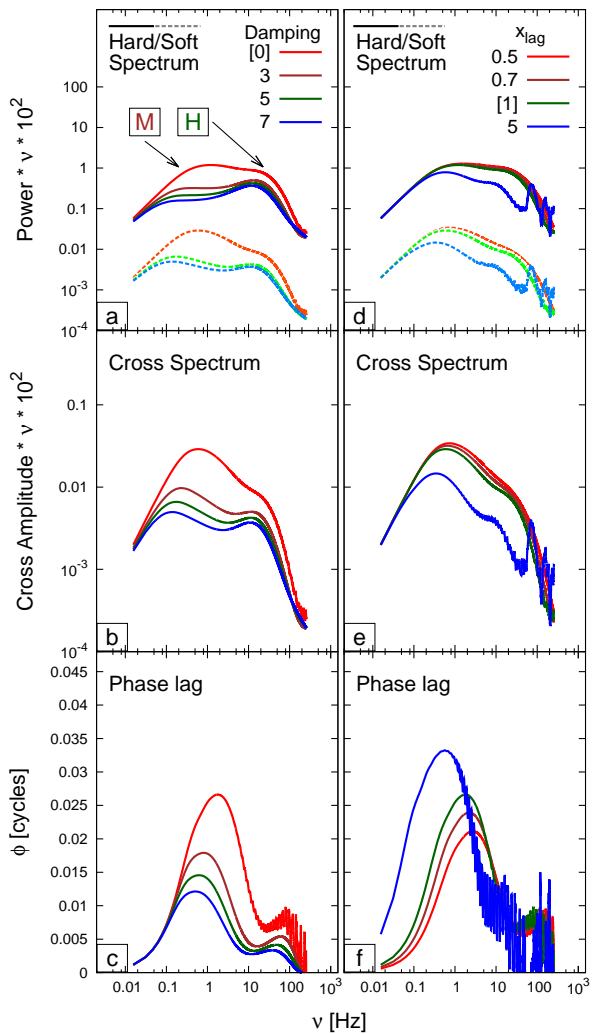}
\caption{Soft (dashed line) and hard (solid line) double hump power spectra, cross-spectra, and phase lags computed varying the damping factor $D$ and $x_{lag}$. The double hump power spectra consist of a main and a high frequency hump (M-H configuration). Numbers in square brackets indicate the parameter value for all the other computations.}
\label{fig:par1}
\end{figure}

The power spectrum of fluctuations stirred up in each ring of the hot flow is a broad Lorentzian with most of the variability power concentrated at the local viscous frequency. This assumption follows the fact that fluctuations with characteristic frequency higher than the local viscous frequency are strongly damped, so they never reach the inner part of the accreting system, where most of the emission comes from (Ingram 2016). This effect was analytically demonstrated by Churazov et al. (2001) for an optically thick geometrically thin disc, and in general it depends on the Green's function of the diffusion equation of the accreting system. \\
In \pf{}, the mass accretion rate at a certain radius $r_\text{n}$ has so far been expressed in the following way (ID12, IK13):
\begin{equation}
\label{eqn1}
\dot{m}(r_\text{n} , t) = \dot{m}_0 (r_\text{n}) \prod^\text{n}_\text{l=1} a_\text{l} (t-\Delta t_\text{ln})
\end{equation} 
where $\dot{m}_0 (r_\text{n})$ is the average mass accretion rate in $r_\text{n}$, $\Delta t_\text{ln}$ is the propagation time from $r_\text{l}$ to $r_\text{n}$, and $a_\text{l}$ is the time series describing the fluctuation amplitude at $r_\text{l}$. \\
Eq. \ref{eqn1} is a special case of the following expression:
\begin{equation}
\label{eqn2}
\dot{m}(r_\text{n} , t) = \dot{m}_0 (r_\text{n}) \prod^\text{n}_\text{l=1} g(r_\text{l},r_\text{n},t) \otimes a_\text{l} (t)
\end{equation}
where the symbol $\otimes$ means convolution and $g(r_\text{l},r_\text{n},t)$ is the Green's function for propagation between $r_\text{l}$ and $r_\text{n}$, i.e. the function describing the diffusion of an infinitely narrow ring of material from $r_\text{l}$ to $r_\text{n}$. Eq. \ref{eqn1} describes inward-only propagation without spreading of the fluctuations, while Eq. \ref{eqn2} describes also inward-only propagation, but this time the ring is allowed to spread and the spreading is regulated by the Green's function $g(r_\text{l},r_\text{n},t)$. When $g(r_\text{l},r_\text{n},t)$ is equal to $\delta(t-\Delta t_\text{ln})$, we obtain Eq. \ref{eqn1}. \\
In order to take into account damping in the updated \pf{} model, we multiply power and cross-spectrum by a factor $e^{-D  \Delta t_\text{ln} \nu}$, where D is a model parameter and $\nu$ is the frequency. This is equivalent to considering a Green's function with the following Fourier transform:
\begin{equation}
\label{eqn3}
G(r_\text{n}, r_\text{l}, \nu) = e^{-D \Delta t_\text{ln}\nu}  e^{-i 2\pi  \Delta t_\text{ln} \nu}
\end{equation}
For every frequency $\nu$, the argument of this Green's function is the phase shift in the fluctuations caused by the time it takes to propagate between the radii $r_\text{l}$ and $r_\text{n}$, and its modulus damps the amplitude of the fluctuations depending to their frequency (high frequencies are more damped) and propagation time (fluctuations propagating from larger radii are more damped). The exact expressions for the power and cross-spectrum including damping are derived in Appendix A. This kind of prescription is analogous to the one used in Arevalo \& Uttley (2006). \\
The left column of Fig. \ref{fig:par1} shows double hump power spectra obtained considering extra variability in the flow (configuration M-H, see previous section) and varying the damping factor $D$. Increasing $D$ suppress the power in both hard and soft band. For the main hump (label M in Fig. \ref{fig:par1}a) the damping is more evident than for the high frequency hump (label H in Fig. \ref{fig:par1}a). This is because fluctuations that propagate longer in the hot flow are more damped. From suppression of variability follows a suppression of phase lags (see in Fig. \ref{fig:par1}c) as more delayed variability is preferentially suppressed. 

\subsection{The propagation delaying factor}
\pf{} so far assumed that the propagation time scale of mass accretion rate fluctuations is equal to the local viscous time scale. In a preliminary analysis of XTE J1550-564 observations, we noticed that the amplitude of the lag predicted by \pf{} was systematically lower than the data. Considering a slower propagation speed would have produced larger lags without significantly modifying the power distribution. For this reason, similarly to Arevalo \& Uttley (2006), we introduced a scaling factor $x_{lag}$ to consider the possibility that fluctuations stirred up in a certain ring can propagate in a time scale not necessary equal (but still proportional) to the local viscous time scale:
\begin{equation}
\Delta t_\text{ln} = x_{lag} \frac{dr}{r} \sum_\text{q=l+1}^\text{n} t_{visc} (r_q) 
\label{eqn4}
\end{equation}
where $\Delta t_\text{ln}$ is the propagation time from $r_\text{l}$ to $r_\text{n}$, $t_{visc}$ is the viscous time scale, and $dr$ is the thickness of the ring. When $x_{lag}$ is larger than 1, the fluctuations propagate slower than the local viscous time scale. Increasing $x_{lag}$ clearly affects the phase lag profile (Fig. \ref{fig:par1}f). As pointed out by Arevalo \& Uttley (2006), varying $x_{lag}$ it is possible to obtain significant lags even if hard and soft emissivity profiles are very similar. However, $x_{lag}$ also affects the power distribution (Fig. \ref{fig:par1}d), introducing complex degeneracy between the parameters regulating the amplitudes of the humps.\\

\section[]{Observations and data analysis}
\label{sec:obs}

We analyzed data from the RXTE Proportional Counter Array (PCA; Jahoda et al. 1996) using 12 pointed observations performed from September 9 (MJD 51065, observation ID 30188-06-01-01) to 1998 September 16 (MJD 51072, observation ID 30188-06-11-00). The observations each contain between $\sim$ 2 and $\sim$ 4 ks of data.\\
Spectral analysis was performed using HEASOFT 6.13: we extracted source and background spectrum from Standard2 data (16 s time resolution). We created a PCA response matrix (necessary for both spectral analysis and \pf{} fitting) for all the analyzed spectra. Each energy spectrum was background subtracted and a systematic error of 0.5\% was applied. We fitted the energy spectrum in the 3-20 keV range using XSPEC 12.8.2 (Arnaud 1996). \\
Following Axelsson et al. (2013), we used the model \textsc{TBABSxGABSx(DISKBB+NTHCOMP+RFXCONVxNTHCOMP)} (Mitsuda et al. 1984; Zdziarski, Johnson \& Magdziarz 1996; $\rm \dot{Z}$ycki, Done \& Smith 1999; Kolehmainen et al. 2001). The obtained results are consistent with the Axelsson et al. (2013) analysis. We extracted source and background light curves in two different energy bands (soft: 1.94-12.99 keV, and hard: 13.36-20.30 keV). From the background subtracted light curves we computed the hardness ratio HR, i.e. the count rate ratio between hard and soft energy band. From spectral analysis we obtained the maximum disc temperature $T_{d,max}$ and we computed the disc fraction in the soft band $x_s$ (for details on the role of HR and $T_{d,max}$ in \pf{} fits, see RIKK16, Appendix A).\\
We used the $\approx$ 125 $\mu$s time resolution Single Bit mode and $\approx$ 16 $\mu$s time resolution Event mode for the Fourier timing analysis. For each observation, we computed Leahy-normalized power spectra in the soft and hard band, and cross-spectra between these two bands, using 256 s data segments and a time resolution of 1/8192 s, giving a frequency resolution of 1/256 $\approx$ 0.004 Hz and a Nyquist frequency of 4096 Hz. Leahy-normalized power and cross-spectra were then averaged, Poisson noise subtracted, and fractional $rms$ normalized (for details about Leahy- and $rms$ normalized cross-spectra, see RIKK16, Sec. 3). We did not apply any background correction in computing the Fourier analysis products. \\
For each observation, by integrating power and cross-spectra, we computed the fractional rms for both hard and soft band, and phase lags between these bands, in three frequency ranges: low (0.004-0.2 Hz), mid (0.5-10 Hz), and high (10-40 Hz).  

\section{Results}
\label{sec:modfit}

\subsection[]{Preliminary analysis}
\begin{figure} 
\center
\includegraphics[scale=0.57,angle=0,trim=0.5cm 0cm 0cm 0cm,clip]{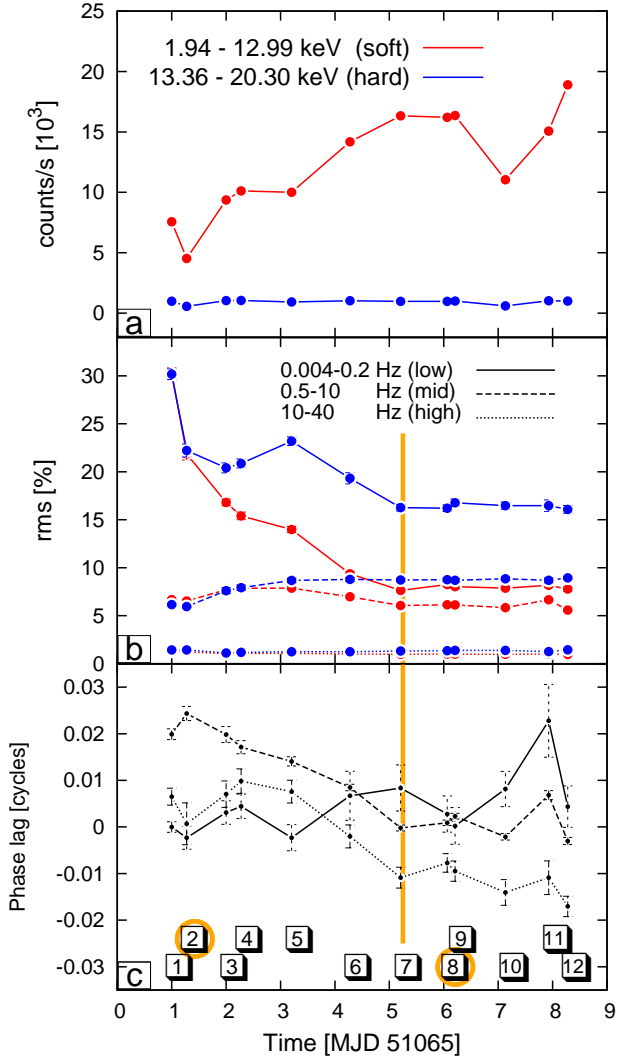}
\caption{Integrated rms ($b$) in the hard (blue points) and soft (red points) band in three frequency ranges: low- (solid line), mid- (dashed line), and high-frequency (dotted line). Panel $c$ shows the phase lags between these two bands integrated over the same frequency ranges. Panel $a$ shows the hard and soft count rate in the full frequency range (0.004-40 Hz). The analyzed observations are labeled from 1 to 12 in panel $c$. The orange vertical bar highlights the transition observed in the characteristics of the source at observation seven. The orange circles indicate the observations fitted with \pf{}.}
\label{fig:rms}
\end{figure}
\begin{figure} 
\center
\includegraphics[scale=0.7,angle=0,trim=0.cm 0cm 0.cm 0cm,clip]{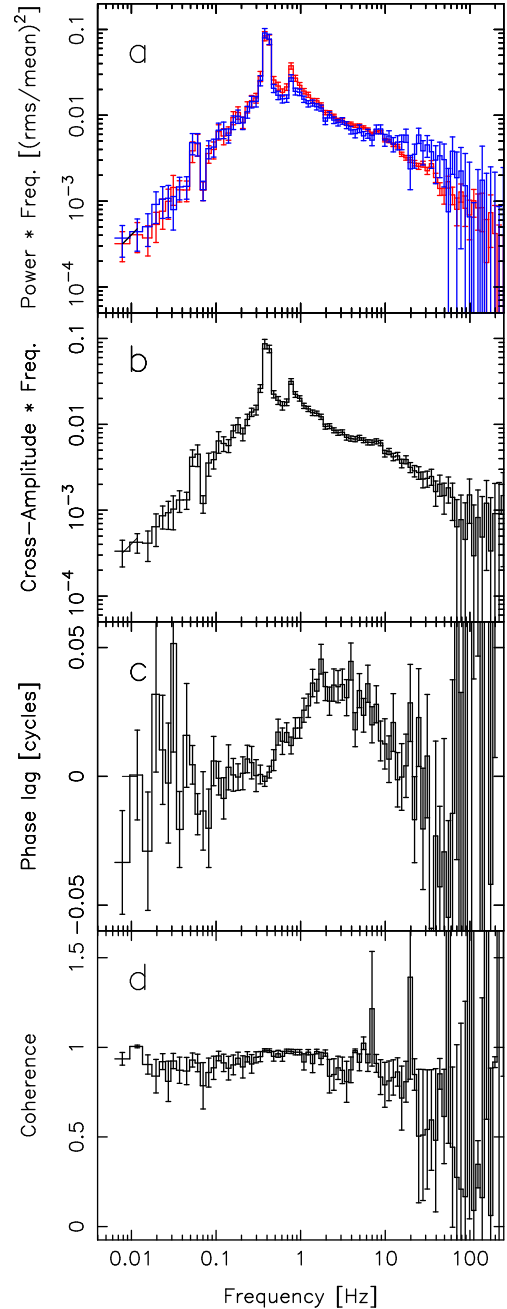}
\caption{Hard (blue line) and soft (red line) power (a), cross-spectrum (b), phase lag (c), and intrinsic coherence (d) of the first observation (30199-06-01-01).}
\label{fig:3plots1}
\end{figure}
\begin{figure} 
\center
\includegraphics[scale=0.7,angle=0,trim=0.cm 0cm 0.cm 0cm,clip]{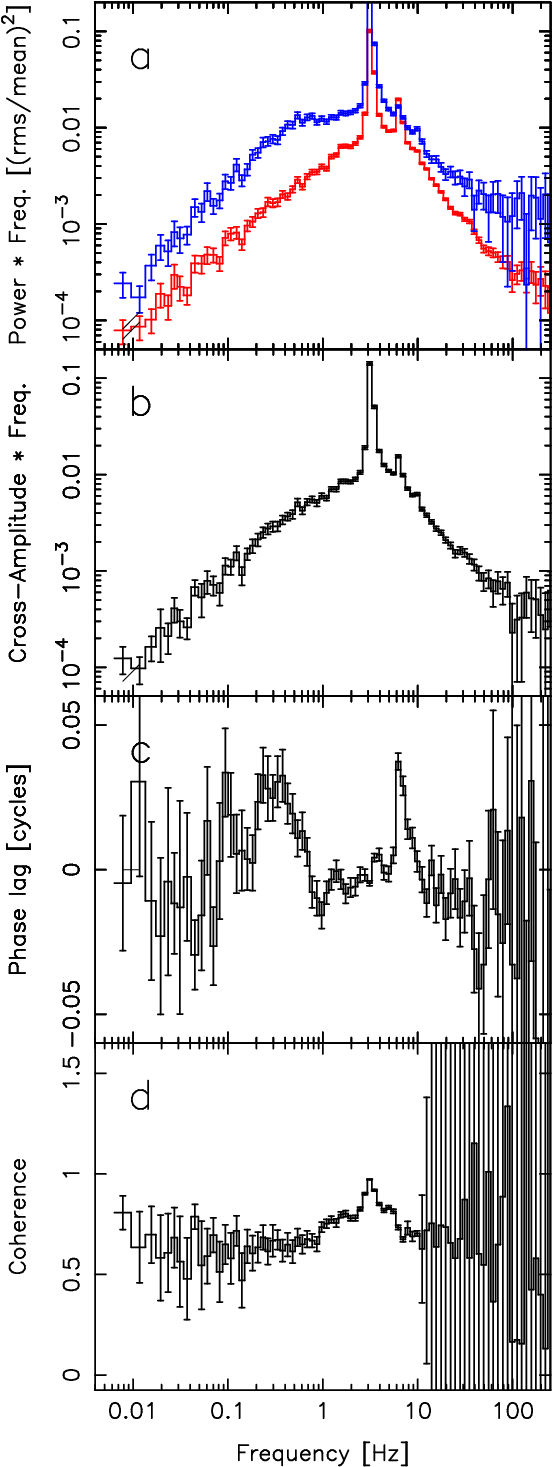}
\caption{Hard (blue line) and soft (red line) power (a), cross-spectrum (b), phase lag (c), and intrinsic coherence (d) of the second observation (30199-06-07-00).}
\label{fig:3plots2}
\end{figure}

Fig. \ref{fig:rms}($b$) shows fractional rms in the hard (blue points) and soft (red points) band for all the analyzed observations. The phase lags between hard and soft band are shown in Fig. \ref{fig:rms}($c$). These quantities are integrated over three frequency ranges: low-, mid-, and high-frequency. Note that the mid-frequency range always contains the QPOs.\\ 
Looking at the evolution of the properties of the source during this first, rising phase, of the outburst, we identify a change in evolution at observation seven (vertical orange line in Fig. \ref{fig:rms}). The main indicators of this change are soft and hard rms in low frequency range, and the phase lag trend in all the frequency ranges.
Up to observation seven, soft and hard rms show a clear decreasing trend in the low-frequency range. From observation seven, hard and soft rms are almost constant in all the frequency ranges. The mid-frequency range phase lag is smoothly decreasing from observation two to seven, leveling off afterward. Up to observation six, low- and high-frequency ranges show smaller phase lags than the mid-frequency range. After an initial correlation from observation one to five, low- and high- frequency ranges show two opposite trends: increasing in the former and decreasing in the latter. From observation seven, the phase lags in the mid-frequency range are close to zero, while the largest lags are now observed in the low-frequency range, and the high-frequency range has negative lags.\\
In order to test \pf{} predictions using the broadest variety of observed characteristics, we selected two observations, before and after the transition respectively. These two observations (orange circles in Fig. \ref{fig:rms}$c$) show very different observational characteristics: in the first selected observation (MJD 51065, observation ID 30188-06-01-01, $\sim$ 2.2 ks, Fig. \ref{fig:3plots1}) soft and hard rms are almost identical at all the frequencies (Fig. \ref{fig:rms}b) and soft and hard power spectra are very similar (Fig. \ref{fig:3plots1}a). This observation is characterized by positive (hard) phase lags that are largest in the mid-frequency range (Fig. \ref{fig:rms}c, dashed line). In the second observation (MJD 51070, observation ID 30188-06-07-00, $\sim$ 3.8 ks, Fig. \ref{fig:3plots2}) the rms in the hard band is larger than in the soft band, especially in the low-frequency range (Fig. \ref{fig:rms}b, solid line), and the hard power exceeds the soft power, principally below 1 Hz (Fig. \ref{fig:3plots2}a). In this observation the largest phase lags are observed in the low-frequency range (Fig. \ref{fig:rms}c, solid line). The two observations clearly differ also in count rate (Fig. \ref{fig:rms}a) and intrinsic coherence (panel $d$ of Fig. \ref{fig:3plots1} and \ref{fig:3plots2}). 

\subsection{\textsc{propfluc} fits}

\begin{figure}
\center
\includegraphics[scale=0.45,angle=0,trim=0cm 1.5cm 0 0cm,clip]{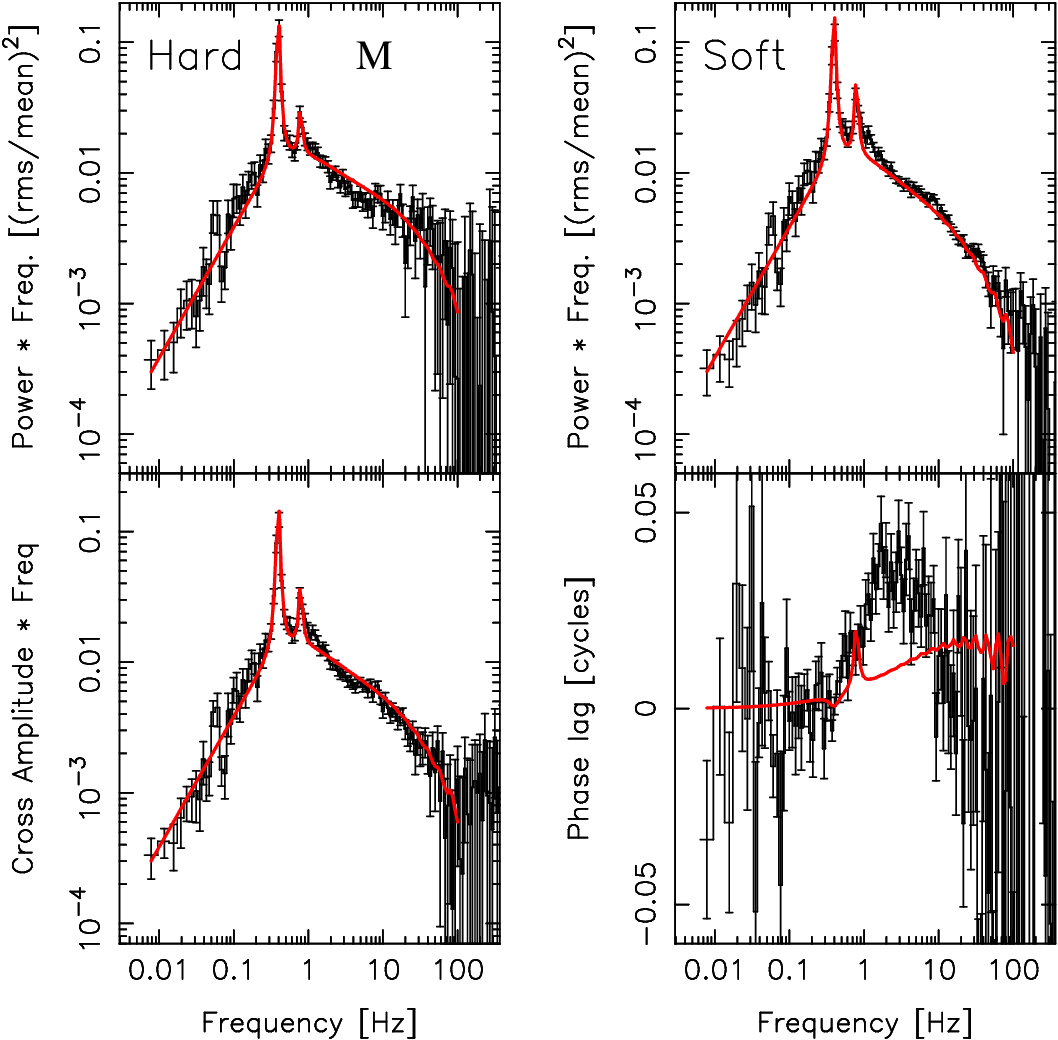}
\vspace{-0.5cm}
\rule[0.525cm]{8cm}{2.pt}
\includegraphics[scale=0.45,angle=0,trim=0cm 1.5cm 0 0cm,clip]{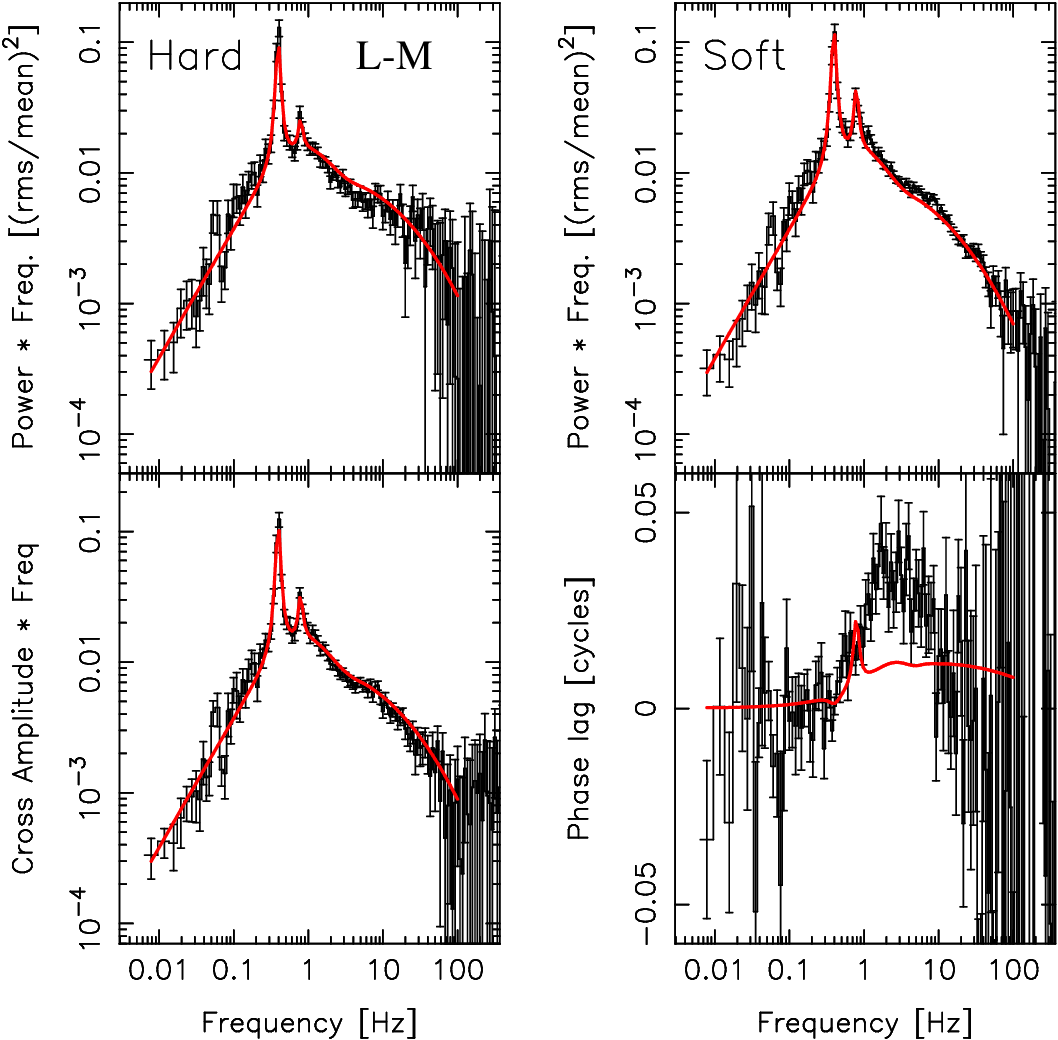}
\vspace{-0.5cm}
\rule[0.525cm]{8cm}{2.pt}
\includegraphics[scale=0.45,angle=0,trim=0cm 0 0 0,clip]{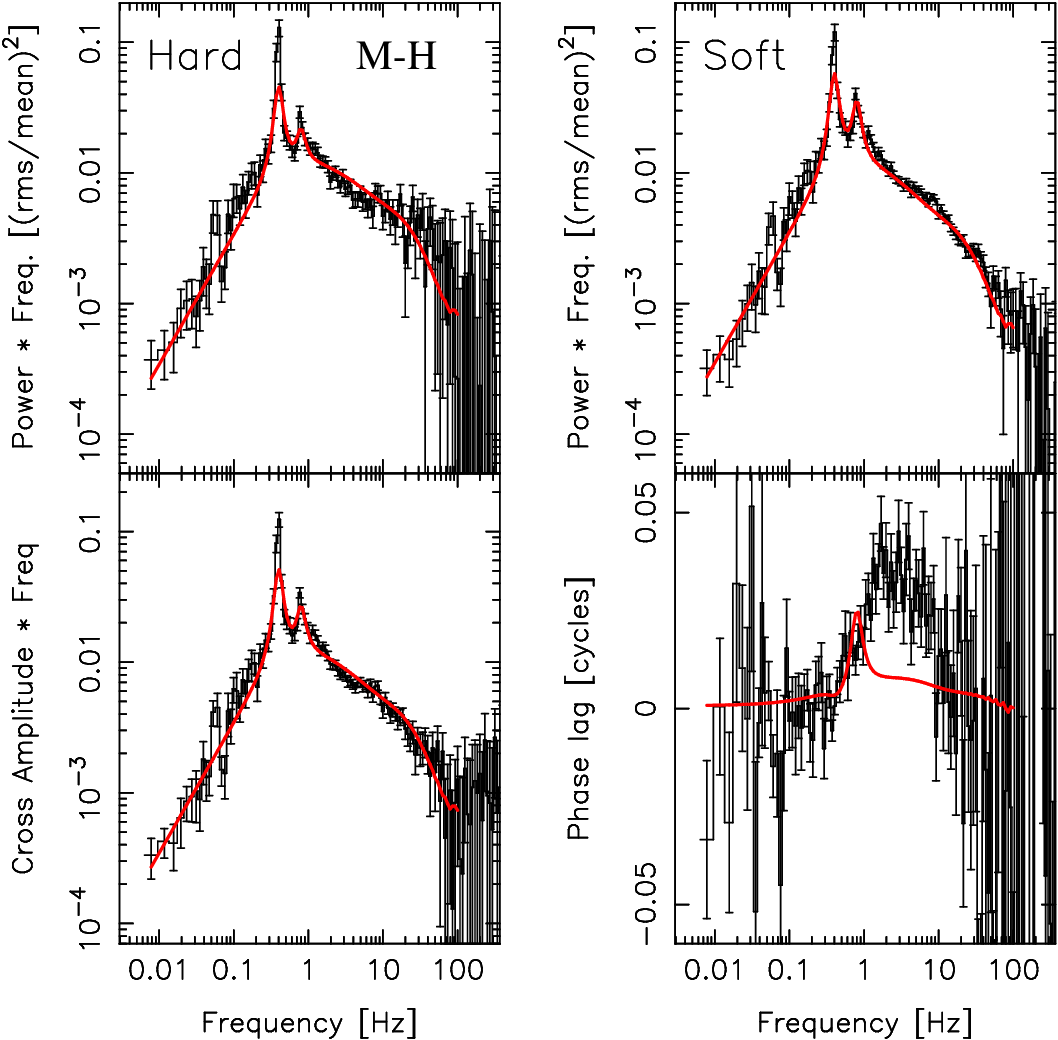}
\caption{Best fit of observation one (solid red line) in three different hump configurations: only main (M), low-frequency and main (L-M), main and high-frequency (M-H).}
\label{fig:obs1}
\end{figure}

\begin{figure}
\center
\includegraphics[scale=0.45,angle=0,trim=0cm 1.5cm 0 0cm,clip]{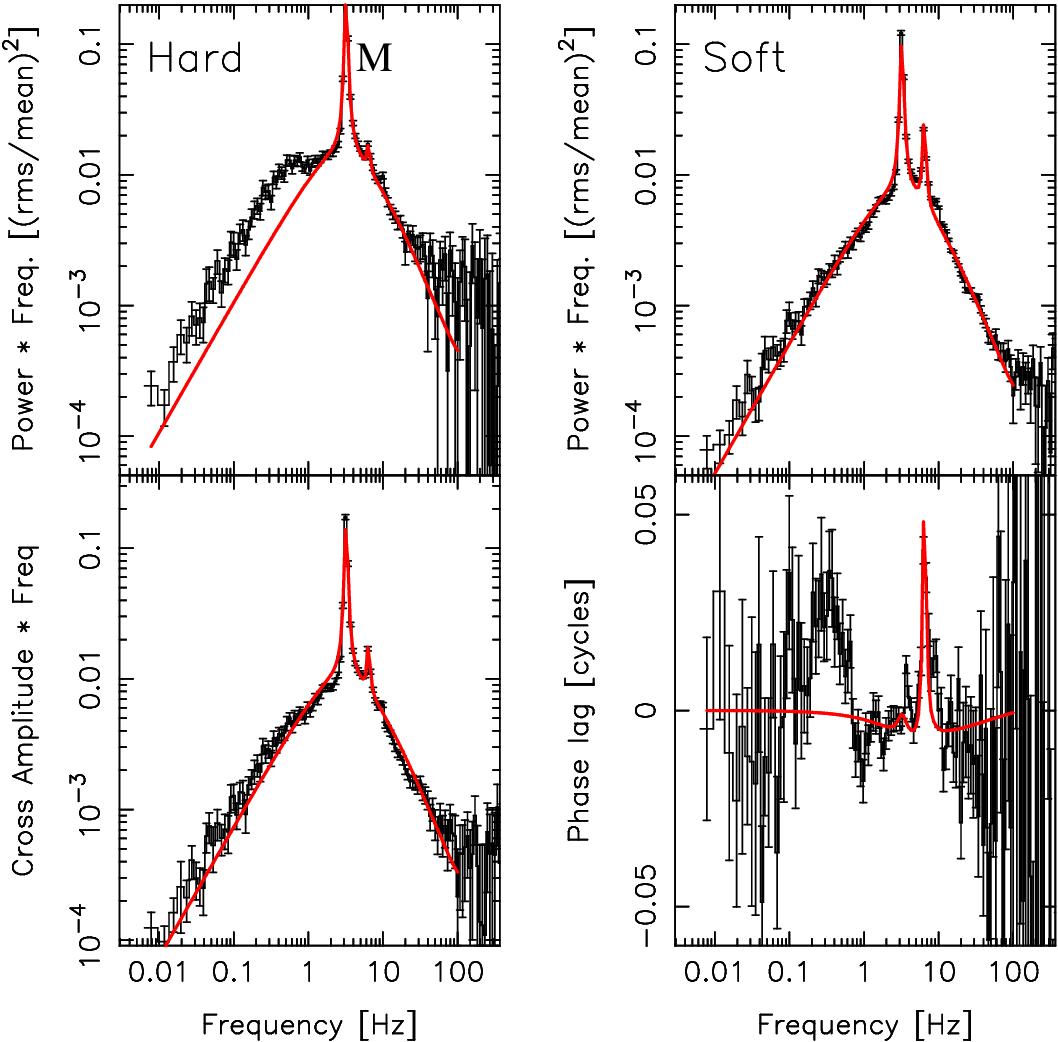}
\vspace{-0.5cm}
\rule[0.525cm]{8cm}{2.pt}
\includegraphics[scale=0.45,angle=0,trim=0cm 1.5cm 0 0,clip]{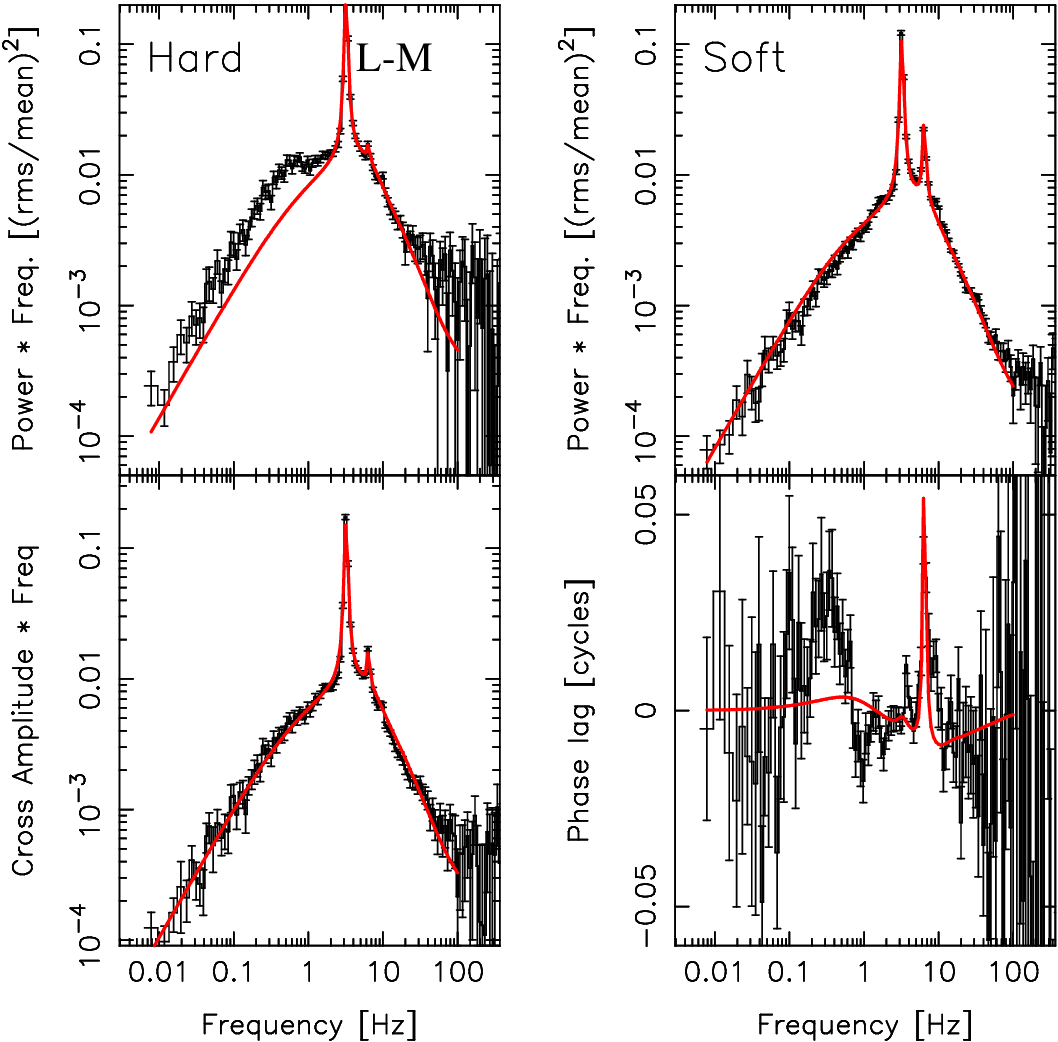}
\vspace{-0.5cm}
\rule[0.525cm]{8cm}{2.pt}
\includegraphics[scale=0.45,angle=0,trim=0cm 0 0 0,clip]{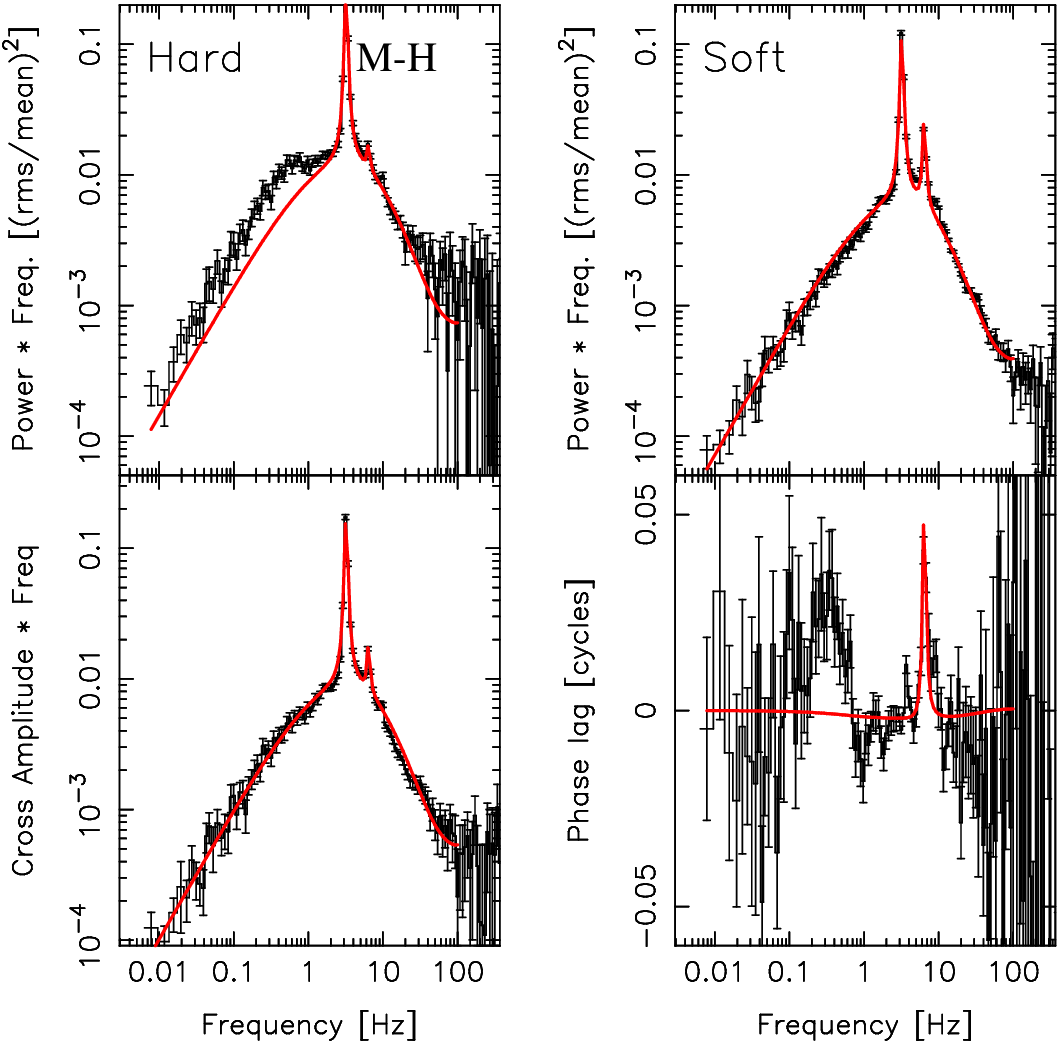}
\caption{Best fit of observation two (solid red line) in three different hump configurations: only main (M), low-frequency and main (L-M), main and high-frequency (M-H).}
\label{fig:obs2}
\end{figure}

We jointly fitted logarithmically binned soft and hard power spectra, and cross-spectra between the two bands in the frequency range 0.004-100 Hz. The resolution is the same for both data and model. A QPO 1$^{st}$ and 2$^{nd}$ harmonic were added to the broad band variability following the prescription described in RIKK16, Appendix B. For a complete description of the model parameters and how these affect cross- and power spectrum, we refer to Rapisarda et al. (2014) and RIKK16. Tab. \ref{tab:par} provides a summary of the model parameters affecting low-, main, and high-frequency hump (the description of QPO parameters is omitted for conciseness). Additionally to what is described in Tab. \ref{tab:par}, we note that the maximum temperature in the disc $T_{d,max}$, the disc fraction in the soft band $x_s$, and the hardness ratio HR (the count ratio ratio between hard and soft band) together set the disc emissivity as a function of radius in the soft and hard band. We obtained all these parameters from spectral analysis and flux measurements. The main parameters related to the QPO (1$^{st}$ + 2$^{nd}$ harmonic rms, quality factor $Q$, and phase lags $\phi$, for a total of five parameters) are free. For all the fits we fixed the number of rings per radial decade ($N_{dec}$ = 35), the hydrogen column density ($n_H$ = 0.6 x 10$^{22}$ cm$^{-2}$, Gierlinski \& Done 2003), the mass ($M$ = 10 $M_{\odot}$), and the dimensionless spin parameter of the BH ($a_*$ = 0.5). For both the observations, we tried to fit the data using four different hump configurations: only main (M, 17 free parameters), main and high-frequency hump (M-H, 18 free parameters), low-frequency and main hump (L-M, 17 free parameters), and the combination of all the three humps (L-M-H, 19 free parameters). In the M configuration, we varied the parameters regulating the surface density profile  ($\lambda$, $\zeta$, $\kappa$), and thereby the radial dependence of the viscous frequency. For all the other configurations we fixed the surface density profile ($\lambda$ = 0.9, $\zeta$ = 0, $\kappa$ = 3). While the number of free parameters in our fit seems quite large, we find that due to the high predictive power of the model, and the highly constraining nature of our formalism, using four independently measured quantities at each Fourier frequency (the power spectrum in the soft and hard band, and the real and imaginary part of the cross-spectrum between these two bands), most of the fit parameters are very well constrained by data of this high quality. Fig. \ref{fig:obs1}-\ref{fig:obs2} show the best fit (solid red curve) in three different configurations (M, L-M, M-H). The best-fit results are reported in Tab. \ref{tab:obs}. The results obtained using the L-M-H configuration are very similar to the L-M configuration, therefore they are omitted in our discussion. The narrow features in the predicted phase lags are associated with the QPO 1$^{st}$ and 2$^{nd}$ harmonic.

\subsubsection{First observation: 30199-06-01-01}
This observation is characterized by a very similar hard and soft power spectral shape, and a phase lag bump between $\approx$ 0.5 and 10 Hz (Fig. \ref{fig:3plots1}).
We could reproduce the power spectral shape with each of the three hump configurations (M, L-M, and M-H, see Fig. \ref{fig:obs1}). The configurations L-M and M-H (double hump power spectrum) better predict the high frequency power. The predicted phase lags clearly disagree with data for all the configurations. In particular, \pf{} underestimates the lag between $\approx$ 0.5 and 10 Hz, so we could not get a statistically acceptable fit (see $\chi^2$ in Tab. \ref{tab:obs}). This is because the emissivity indices necessary to fit the power are too similar to produce the observed lags (see $\Delta \gamma$ in Tab. \ref{tab:obs}).

\subsubsection{Second observation: 30199-06-07-00}
This observation is characterized by a different hard and soft power spectral shape, in particular the fractional rms in the hard band is higher than in the soft band at all the frequencies and the hard power spectrum shows a broad shoulder at $\approx$ 0.7 Hz absent from the soft power spectrum (Fig. \ref{fig:3plots2}). The main feature in the phase lag profile is a bump between $\approx$ 0.1 and 1 Hz. Fitting the data with the all the available configurations, we could not reproduce the difference between soft and hard power spectral shape. In particular, \pf{} always underestimates the hard power below $\approx$ 0.7 Hz (Fig. \ref{fig:obs2}). Furthermore, the best-fit emissivity index in the soft band is always larger than in the hard band. This suggests a rather unphysical scenario where the softer emission comes from the inner part of the flow. The phase lags are also underestimated between $\approx$ 0.1 and 0.7 Hz. All these discrepancies lead, again, to a statistically unacceptable fit (see $\chi^2$ in Tab. \ref{tab:obs}).   

\section{Discussion}
\label{sec:dis}
We tested the propagating fluctuations model on two high s/n RXTE observations of the BH XTE J1550-564 during the rising phase of its 1998-1999 outburst, relaxing several of the constrains on the generation and propagation of the fluctuations by introducing additional free parameters. We jointly fitted power density spectra in two energy bands and the complex cross-spectrum between the two bands. This is the most stringent test of the model to date requiring the model to correctly predict amplitude, time lags, and coherence of the variability in and between the two energy bands chosen: our formalism uses all the information contained in the spectral variability that can be extracted by Fourier analysis in two energy bands.\\  

\subsection{Fit to XTE J1550-564 observations}
We fitted the two selected observations considering all the possible hump configurations: M, L-M, M-H, and L-M-H. Additionally to previous applications of \pf{}  (ID12; IK13; Rapisarda et al. 2014; RIKK16), we considered damping, different fluctuation propagating speeds, and different hump configurations. Even using these new features we found that the model cannot explain the data: the $\chi^2$ is not acceptable and, even more importantly, there are qualitative differences between observed power/cross-spectra and model predictions. As described below, these discrepancies appear to be generic to the propagating fluctuations idea. \\
\begin{figure}
\center
\includegraphics[scale=0.45,angle=0,trim=0cm 0 0 0,clip]{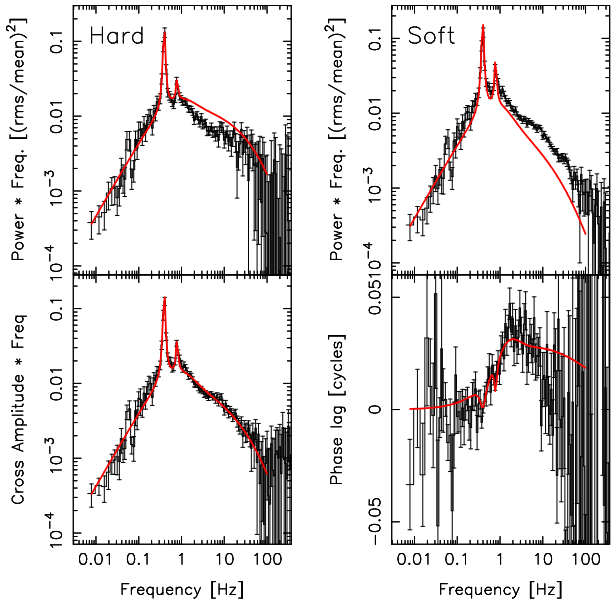}
\caption{Best fit of observation one (red line) obtained using the M (single hump) configuration and fitting only the cross-spectrum}
\label{fig:propC}
\end{figure}
The first observation shows hard lags (the bump between $\approx$ 0.7 and 10 Hz) in combination with almost identical hard and soft power spectra up to $\approx$ 10 Hz. Propagating fluctuation models assume that different timescale variability is generated at different radii and that mass accretion rate fluctuations propagate from larger to smaller radii. In such scenarios, the positive (hard) phase lags occur because the hard band has a steeper emissivity profile than the soft band, as previously concluded by, e.g., Kotov et al. (2001) and Arevalo \& Uttley (2006). Therefore, the hard power spectrum is more weighted to the faster characteristic frequencies of the inner regions than the soft band. When we use \pf{} to quantitatively confront these model ideas with observations, we find that in the first observation the similarity between soft and hard power spectra up to $\approx$ 10 Hz requires emissivity profiles for the two bands that are very close to each other (see $\Delta \gamma$ in Tab. \ref{tab:obs}), which, as expected from the generic argument above, leads to predicted phase lags close to zero, in evident contradiction to the data. We note that fitting \emph{only} the cross-spectrum, \pf{} can reproduce the phase lags of the first observation, but the best-fit emissivity indices do not reproduce the power spectral shape observed in the soft and hard band (see Fig. \ref{fig:propC}). When we jointly fit power and cross-spectrum, we obtain a better $\chi^2$ reproducing the shape of the power spectrum instead of the phase lags. This is because the error bars of the power are smaller than the phase lags. \\ 
The second observation shows a difference between hard and soft power spectra; in particular the hard power spectrum is characterized by a low frequency feature (around 0.7 Hz) that is absent from both the soft power and the cross-spectrum (see Fig. \ref{fig:3plots2}a and b). This suggests that the variability associated with this feature is uncorrelated with the variability in the soft band, as indeed confirmed by the small coherence at low frequency (Fig. \ref{fig:3plots2}c). We do not expect this behavior if this slow variability originates from farthest from the BH and the hard band has a steeper emissivity than the soft band.\\
The observations, when considered quantitatively in the context of a propagation model, seem to have characteristics that are generically at variance with propagating fluctuation model expectations, and indeed our \pf{} fit attempts confirm this. In a geometry in which the spectral hardness, variability timescale and propagation speed are all single valued functions of radius, the observed data cannot be explained by a model in which lags are produced by fluctuations propagating towards the black hole. It is possible that a more complex geometry could explain the data (such as considering an overlap region between disc and inner flow), but detailed modelling is required to make any robust predictions. \\
In the following sections, we discuss different factors that could explain the discrepancies between \pf{} predictions and data.

\subsection{Possible explanations}

\subsubsection{QPO lags}
In the observations we analyzed (LHS of XTE J1550-564), the QPO always dominates the rms and its frequency increases from $\approx$ 0.3 to 10 Hz. The presence of the phase lag bump between $\approx$ 1 and 10 Hz prevents us to clearly identify the lag associated with the QPO and its harmonics in the first part of the outburst (up to observation 6, see Fig. \ref{fig:rms}). \\
In \pf{} the QPO is modeled with an \textit{ad hoc} prescription producing features in the phase lag versus frequency spectrum similar in width to those in the power spectrum (see RIKK16, Appendix B). While this simple solution allowed to fit data characterized by strong QPOs (Rapisarda et al. 2014; RIKK16), this means we do not take into account possible more complex QPO characteristics. For example, we can not exclude that the QPO has broad higher harmonics contributing to both the power and the phase lag. Alternatively, the presence of the QPO might be part of a physical process that also influences the broad band noise. Perhaps it is possible to obtain clues about the role of QPOs in the phase lag spectrum by comparing to sources that do not show strong QPOs, but similar broad band noise and energy spectrum (Rapisarda et al. in prep.). In this context, the propagating fluctuations model could also be applied to cataclysmic variables (Scaringi 2014). Indeed, the accretion flow in these sources produces variability similar to BHB variability (e.g. Scaringi et al. 2012; Scaringi et al. 2015), but it is not affected by strong general relativistic effects. 

\subsubsection{Reflection}
\begin{figure} 
\center
\includegraphics[scale=0.55,angle=0,trim=0.cm 0cm 0.cm 0cm,clip]{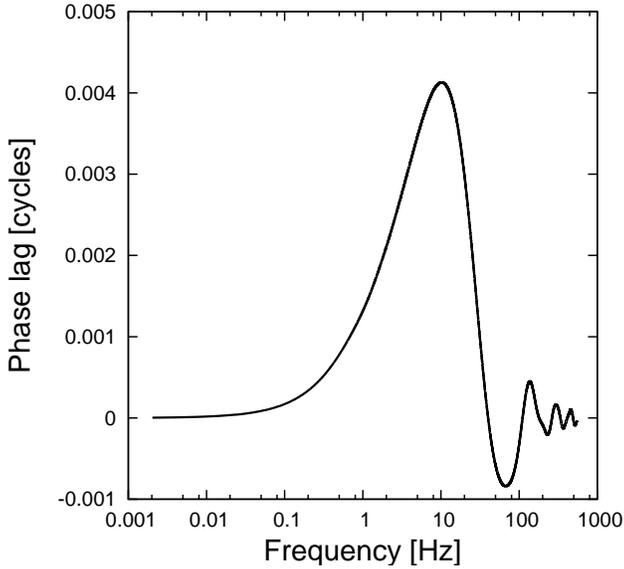}
\caption{Phase lags resulting from reflection of X-ray emitted by an isotropic stationary source located 200 $R_g$ above the BH. The amplitude of the lags is a factor 10 smaller than the data.}
\label{fig:refl}
\end{figure}
\begin{figure} 
\center
\includegraphics[scale=0.55,angle=0,trim=0.cm 0cm 0.cm 0cm,clip]{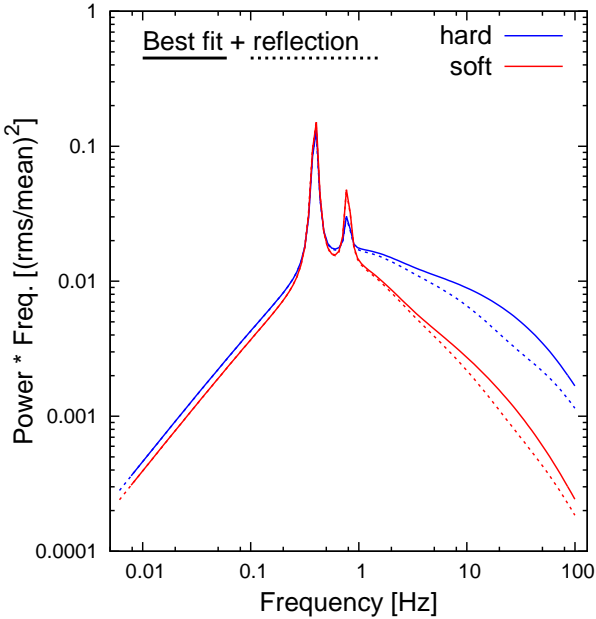}
\caption{Power spectra resulting from including reflection to propagating fluctuations. Due to light travel effects, the reflected component dilutes the emission suppressing mainly the hard power at high frequencies.}
\label{fig:reflII}
\end{figure}
The phase lags predicted by \pf{} are only due to the propagation of mass accretion rate fluctuations. However, it is well known that reflection contributes significantly to the spectrum, and this is true also for our selected observations (Axelsson et al. 2013). Photons from the hot flow are reflected by the disc, so that the reflected emission is delayed depending on the finite light travel time from the hot flow to the disc and on the size of the reflector (the disc). For the energy bands considered in our analysis, we expect more reflection in the hard band than in the soft band (since the Compton hump is harder than the continuum), so that reflection can contribute hard lags. Because of the finite size of the disc, the high frequency variability of the reflected component is suppressed (e.g. Gilfanov et al. 2000), so that the reflected emission affects both the way the soft and hard power are distributed between the frequencies, and the amplitude of the (hard) phase lags. In the first observation, the phase lag bump observed between $\approx$ 1 and 10 Hz cannot be produced by propagating fluctuations because of the similarity between the hard and soft power spectra. However, frequency dependent phase lags will be produced by hard photons emitted by the hot flow and reflected from the disc. \\
We tested the contribution of reflection by calculating a transfer function from a lamppost model formalism (X-rays are emitted from an isotropic stationary point some height $h$ above the BH), ignoring light bending (e.g. Campana \& Stella 1995). We assume the inner boundary of the reflector, $r_{in, refl}$, equal to the best-fit truncation radius for the first observation, \ro{} = 42.3, $i = 70$ degrees. Our spectral fitting (see Sec. \ref{sec:obs}) indicates that reflection accounts for $\approx$ 16\% of the soft band flux and $\approx$ 23\% of the hard band flux. We calculate the hard lags that would result purely from reflection. Fig. \ref{fig:refl} shows that, even assuming a very large h = 200 \rg{}, the magnitude of the lag ($\approx$ 0.003 cycles) is only a small fraction of the observed lag ($\approx$ 0.04 cycles), and the lag peaks at a higher frequency than observed (we assume a 10 $M_{\odot}$ BH; increasing the mass would lower the frequency where the lag peaks). We note that it is possible to reproduce a lag of the required amplitude with reflection, but for rather implausible parameters. This requires considering the hard emission to come from $\approx$ 1 \rg{} and the soft emission to come from $\approx$ 200 \rg{}, which is unlikely.\\
Fig. \ref{fig:propC} shows that \pf{} can reproduce the phase lags observed in the first observation at the expense of a correct prediction of the power spectral shape in the soft and hard band. As mentioned at the beginning of this section, reflection can affect the shape of the power spectrum in the two bands. In particular, we want to explore the possibility that the \emph{intrinsic} soft and hard power spectra are different from one another (allowing to produce the observed hard lag), but, because of reflection, their shapes are modified to be similar to one another, as observed in the data. In order to test the effect of reflection on the power distribution, we computed power spectra from the fit parameters of Fig. \ref{fig:propC} including reflection and using the same setting adopted above (lamppost model, \ro{} = 42.3, $i = 70$, h = 200 \rg{}, see Appendix B for details). The power spectra obtained including reflection are shown in Fig. \ref{fig:reflII}. Including reflection suppresses the power at high frequency, mainly in the hard band (this is because the hard band contains more reflection than the soft band). Although including reflection reduces the difference between hard and soft power spectrum, the model prediction is still far from the observed power spectral shapes (almost identical power distribution in both energy bands). Therefore, reflection, either purely or combined with propagating fluctuations, is unlikely to explain the discrepancies observed between data and model predictions. We note that Walton et al. (2013) reported on a variable AGN with almost no reflection feature, so this would provide a very good test for the propagation model.\\
We note that a fraction of the hard X-ray photons coming from the hot flow and intercepted by the disc can be reprocessed. However, independently of the reprocessing time scale, this process will contribute a \emph{soft} phase lag associated with the high frequency variability, and therefore cannot explain the hard lags we observe.

\subsubsection{Inwards and outwards propagation}
As described in Sec. \ref{sec:up}, \pf{} assumes a simplified diffusion mechanism, with fluctuations propagating only towards the BH and without spreading of the fluctuations during propagation. A more realistic approach would require considering a Green's function closer to proper diffusion, i.e. accounting for inwards and outwards propagation. Outward propagation could introduce soft phase lags leading to unintuitive results. This is beyond the scope of this study, but we highlight the importance of exploring fundamental principles of the propagation models as currently understood to check if the discrepancies we found in these XTE J1550-564 observations can be naturally addressed in a more physical description of the diffusion process (Mushtukov et al. in prep).

\subsubsection{Non-linear spectral variations}
\pf{} assumes that the flux in a certain energy band is a linear combination of the mass accretion rate weighted fluxes of each ring (IK13). If the shape of the energy spectrum from each ring is not constant but varies with mass accretion rate, this approximation is not valid anymore. In a more realistic scenario, we expect the blackbody disc temperature and the Compton power law index to vary with mass accretion rate. If these variations are small enough, we could still represent the flux as a linear combination of the mass accretion rate at each ring linearizing the variations through Taylor expansion. So, in this case, we would not expect significant differences from the actual \pf{} predictions. Larger variations would lead to deviations from the power spectral shape and the phase lag profile predicted by a linear model like \textsc{propfluc}. Exploring the characteristics of these deviations requires further quantitative exploration of our model. We plan to investigate the magnitude of these deviations in a future analysis.

\section{Conclusions}
\label{sec:con}
We used the propagating mass accretion rate fluctuations model \pf{} to fit two observations in the LHS of XTE J1550-564. In our fit, we accounted for damping and different fluctuation propagation speeds. The model failed to reproduce the data both quantitatively (unacceptable reduced $\chi^2$ values) and qualitatively. In the first observation, this could still be due to the simplistic QPO prescription adopted in \pf{}. However, in the second observation, the discrepancies between data and model predictions are a generic issue for propagation models and not specific to the \pf{} implementation. This result encourages a deeper investigation of the fundamental properties of the propagation model, with particular attention to the physical mechanism producing the QPO and how it affects the broad band noise.\\
\\
\textbf{ACKNOLEDGEMENTS}\\
We thank the referee for his useful comments that helped to improve the manuscript. The authors acknowledge support from the Netherlands Organization for Scientific Research (NWO).

\begin{appendix}
\section{Including damping}
\label{app:damp}

Eq. \ref{eqn2} describes the mass accretion rate at the
$n^{\rm th}$ ring for a general Green's function assuming that fluctuations propagate only inwards. If we inject a
narrow ($\delta-$function) ring of mass into the disc, the Green's
function describes how that ring spreads out and eventually accretes
onto the BH. Considering the Green's function  of IK13 ($g(r_{n=1},r_\text{n},t) = \delta(t-\Delta
t_{l,n})$), this ring would remain a narrow ring and simply move
towards the BH at a velocity $r \cdot \nu_v(r)$. In a more realistic scenario, the
ring will spread: some material will move towards the BH slower or
faster than the average rate, and in general some material can even
spread outwards from the starting point. Here, we generalise the IK13
formalism by considering a more general Green's function. We still
assume only inwards propagation (since this vastly simplifies the Green's function solutions: Mushtukov, Ingram \& van der Klis in prep.), but the form
for the Green's function given in Equation (3) in the text allows for
spreading of the ring.\\
\begin{figure} 
\center
\includegraphics[scale=0.45,angle=0,trim=0.5cm 0cm 0cm 0cm,clip]{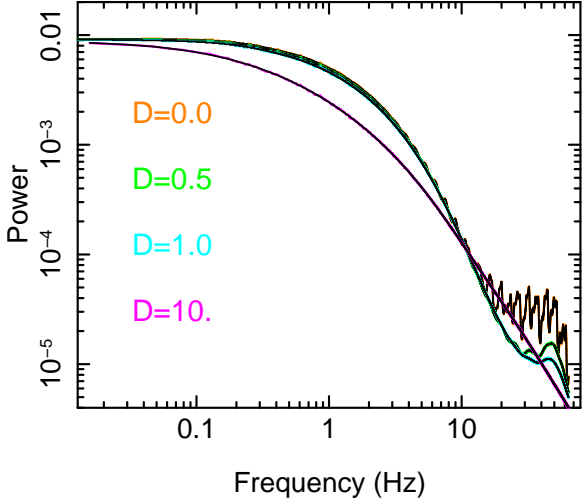}
\caption{Single hump power spectrum (M configuration) predicted by the model. Different colours represent different values of the damping factor $D$. The colored lines are the average of 10000 iterations obtained using the Timmer \& Koening (1995) prescription and the black lines are the analytic calculations.}
\label{fig:carlo}
\end{figure}
Equation (2) implicitly assumes that the Green's function has the
property:
\begin{equation}
g(r_\text{l},r_\text{n},t) = g(r_\text{l},r_q,t) \otimes g(r_q,r_\text{n},t),
\end{equation}
or equivalently in Fourier space:
\begin{equation}
G(r_\text{l},r_\text{n},\nu) = G(r_\text{l},r_q,\nu) G(r_q,r_\text{n},\nu).
\end{equation}
where ($r_\text{l} > r_q > r_\text{n}$). For a function describing propagation, it is rather unphysical to not have this property, since a fluctuation propagating from $r_1$ to $r_3$ must equal the same fluctuation propagating first from $r_1$ to $r_2$, and
then from $r_2$ to $r_3$. The IK13 Green's function, $G(r_\text{l},r_\text{n},\nu) =
\exp(i 2\pi \Delta t_{l,n} \nu)$, has this property since $\Delta
t_{l,n}= \Delta t_{l,q} + \Delta t_{q,n}$. The Green's function
implemented here (Eq. \ref{eqn3}) also has this
property. Because this property holds, the same arguments can be used
as in IK13 (Appendix B therein) to show that the power spectrum of the
mass accretion rate at $r_\text{n}$ is
\begin{equation}
|\dot{M}(r_\text{n},\nu)|^2 = \coprod_{l=1}^n |G(r_\text{l},r_\text{n},\nu)|^2 |A_\text{l}(\nu)|^2.
\label{eqn:Pmdot}
\end{equation}
Similarly, the cross-spectrum between $\dot{M}(r_\text{n},\nu)$ and
$\dot{M}(r_\text{l},\nu)$ is
\begin{equation}
\dot{M}(r_\text{n},\nu)\dot{M}^*(r_\text{l},\nu) = \Lambda_{l,n} G(r_\text{l},r_\text{n},\nu)
|\dot{M}(r_\text{l},\nu)|^2,
\label{eqn:mcross}
\end{equation}
where $\Lambda_{l,n}$ is the product of the mean mass accretion rate
in each radius (following IK13). Since we assume the same mass
accretion rate in each ring, we can always set $\Lambda_{l,n}=1$.\\
The power spectrum of the ``hard band'' flux becomes
\begin{eqnarray}
& &P(\nu) = \sum_{n=1}^{\mathcal{N}} \bigg[ 
h_\text{n}^2 |\dot{M}(r_\text{n},\nu)|^2  \nonumber \\
& & + 2\sum_{l=1}^{n-1} h_\text{l} h_\text{n} \Lambda_\text{ln} |G(r_\text{l},r_\text{n},\nu)| \cos\Phi_{l,n}(\nu) |\dot{M}(r_\text{l},\nu)|^2 \bigg],
\label{eqn:Pf}
\end{eqnarray}
where $\Phi_{l,n}(\nu) \equiv \arg[G(r_\text{l},r_\text{n},\nu)]$. For our Green's
function (and also that of IK13), $\Phi_{l,n}(\nu) = 2\pi \Delta
t_{l,n} \nu$. Fig. \ref{fig:carlo} shows that this calculation (black lines) gives
the same result as using the average of 10000 realisations obatined using the Timmer \& Koenig (1995) prescription for a range of values of the damping factor $D$ (coloured
lines). This can be extended to the cross-spectrum between the ``hard''
and ``soft'' band, giving
\begin{eqnarray}
C(\nu) & =& \sum_{n=1}^{\mathcal{N}} \bigg[ 
h_\text{n} s_\text{n} |\dot{M}(r_\text{n},\nu)|^2  \\
&+ &\sum_{l=1}^{n-1} (h_\text{l} s_\text{n} e^{i\Phi_{l,n}(\nu)}+h_\text{n} s_\text{l}
  e^{-i\Phi_{l,n}(\nu)} ) \nonumber\\ 
&\times& \Lambda_\text{ln} |G(r_\text{l},r_\text{n},\nu)| |\dot{M}(r_\text{l},\nu)|^2
\bigg]. \nonumber
\end{eqnarray}

\section{Including reflection}
\label{app:refl}
We can represent the hard (soft) band flux as the sum of a direct and reflected component:
\begin{eqnarray}
h(t) & = & d_h(t) + x_{r,h} [d_h(t) \otimes \tau(t)]
\end{eqnarray}
where $d_h(t)$ ($d_s(t)$) is the direct component in the hard (soft) band, $x_{r,h}$ ($x_{r,s}$) is the reflection fraction in the hard (soft) band, $\tau(t)$ is the impulse response function or transfer function, and $\otimes$ means convolution. We calculated the transfer function from a lamppost model formalism (X-rays are emitted from an isotropic stationary point at some height $h$ above the BH) and ignoring light bending (e.g. Campana \& Stella 1995). Normalizing $d_h(t)$ ($d_s(t)$) to have a mean of unity, the mean of $h(t)$ is $1+x_{r,h}$ (the mean of $s(t)$ is $1+x_{r,s}$).\\
Applying the Fourier transform, we obtain:
\begin{eqnarray}
H(\nu) & = & D_h(\nu) [1 + x_{r,h} \times T(\nu)]
\label{fourier}
\end{eqnarray}
and the cross-spectrum between soft and hard band is:
\begin{eqnarray}
S^{*}(\nu)H(\nu) = D_s^{*}(\nu) D_h(\nu) [1 + x_{r,s} T^{*}(\nu)][1 + x_{r,h} T(\nu)]
\label{cross}
\end{eqnarray}
The power spectrum in the hard (soft) band is:
\begin{eqnarray}
|H(\nu)|^2 = |D_h(\nu)|^2[1+ x_{r,h}^2 |T(\nu)|^2 + 2 x_{r,h} \Re{T(\nu)}]
\end{eqnarray}
Representing the power spectrum in the fractional rms normalization we obtain:
\begin{eqnarray}
P_h(\nu) & = & |H(\nu)|^2 / (1+x_{r,h})^2 \\
         & = & \frac{|D_h(\nu)|^2[1+ x_{r,h}^2 |T(\nu)|^2 + 2 x_{r,h} \Re{T(\nu)}]}{(1+x_{r,h})^2} \nonumber
\end{eqnarray}
The presence of a stable reflected component affects both the shape and the normalization of the power spectrum.

\end{appendix}

\begin{table*}
\caption{Short description of \pf{} parameters.}
\label{tab:par}
\begin{tabular}{c|p{10cm}}

\multicolumn{2}{c}{\Large{Main hump parameters}} \\
\hline\hline
$\Sigma_0$ & Surface density constant. It is a multiplicative factor for the surface density in the hot flow and it regulates the characteristic frequency of the main hump M.\\ 
\arrayrulecolor{black}\hline
$F_{var}$ & Fractional variability per radial decade in the hot flow. It regulates the amplitude of the main hump.\\
\hline
$\lambda$ & Power law index determining the surface density profile in the hot flow for radii much smaller than the bending wave radius.\\
\hline
$\zeta$ & Power law index determining the surface density profile in the hot flow for radii much larger than the bending wave radius.\\
\hline
$\kappa$ & Index regulating the sharpness of the transition between $\lambda$ and
$\zeta$ regimes at the bending wave radius.\\
\hline
$r_{in}$ & Inner radius of the hot flow.\\
\hline
$r_{bw}$ & Bending wave radius. $r_{bw}$ is in between $r_{in}$ and $r_o$. \\
\hline
$r_o$ & Truncation radius. It sets the outer edge of the hot flow and it is the main parameter affecting the QPO frequency. \\
\hline
$\gamma_s$ & Emissivity profile index of the hot flow in the soft band. \\
\hline
$\gamma_h$ & Emissivity profile index of the hot flow in the hard band. \\
\hline
\hline
\\

\multicolumn{2}{c}{\Large{Low-frequency hump parameters}} \\[0.1cm]

\hline
\hline
$r_o$ & Truncation radius. It sets the inner edge of the disc.\\
\hline
$r_d$ & Outer edge of the varying disc.\\
\hline
$\nu_{d,max}$ & Maximum viscous frequency in the disc. It regulates the characteristic frequency of the low-frequency hump L.\\
\hline
$N_{var}$ & Amount of variability produced in the disc as a fraction of $F_{var}$. It regulates the amplitude of the low-frequency hump. When $N_{var}$ = 0, no variability is produced in the disc. When $N_{var}$ = 1, the amount of variability produced at the truncation radius in the disc is equal to $F_{var}$. \\
\hline
$\Delta d$ & Radial extension of the varying disc. It affects both the characteristic frequency and the amplitude of the low-frequency hump H.\\
\hline
$T_{d,max}$ & Maximum temperature in the disc. \\
\hline
$x_s$ & Fraction of disc photons emitted in the soft band. It affects the amplitude of the phase lag associated with the variability propagating from the disc. \\
\hline
\hline
\\ 
\multicolumn{2}{c}{\Large{High-frequency hump parameters}} \\[0.1cm]

\hline 
\hline
$r_{bw}$ & Bending wave radius. It is the radius in the hot flow where the extra variability peaks. \\
\hline
$N_{extra}$ & Amplitude of the extra variability in the flow.\\
\hline
$\Delta r$ & Width of the high-frequency hump.\\
\hline
\hline

\\
\multicolumn{2}{c}{\Large{Global parameters}} \\[0.1cm]
\hline\hline 
$M_{BH}$ & Mass of the BH.\\ 
\hline
$a_{*}$ & Dimentionless spin parameter of the BH.\\
\hline 
$n_H$ & hydrogen column density in units of $10^{21}$ [atoms/cm$^2$]. \\
\hline
\hline

\\
\multicolumn{2}{c}{\Large{Other information}} \\[0.1cm]
\hline\hline 
$HR$ & Hardness ratio: count ratio between hard and soft band.  \\
\hline 
$N_{dec}$ & Number of rings per radial decade in the hot flow and in the disc. It sets the resolution of the model. \\
\hline
\hline

\end{tabular}
\end{table*}

\begin{table*}
\setlength\extrarowheight{5pt}
\caption{\pf{} best-fit parameters of the first and second observation selected in our study (obs. ID 30188-06-01-01 and ID 30188-06-07-00, respectively). Each observation has three columns corresponding to three different hump configurations: only main hump (M), main and high-frequency hump (M-H), and low-frequency and main hump (L-M). The subscripts $s$ and $h$ correspond to soft and hard band respectively. The symbol $\sim$ means that the parameter is fixed at the value in the previous column, the symbol - means that the parameter value is irrelevant for that particular configuration.}
\label{tab:obs}

\begin{tabular}{c|ccc||ccc|}

\hline

Observation & \multicolumn{3}{c||}{First: 30188-06-01-01} & \multicolumn{3}{c|}{Second: 30188-06-07-00}\\
\hline

Humps & M & L-M & M-H & M & L-M & M-H\\
\hline\hline

$\Sigma_0$ & 19.61 & 0.26 & 3.99 & 1.71 & 1.28 & 6.38\\
$F_{var} [\%]$ & 27.18 & 50.45 & 33.27 & 106.28 & 114.47 & 36.05\\
$\zeta$ & 1.9 & 0.0 & 0.0 & 0.0 & 0.0 & 0.0\\
$\lambda$ &2.7 & 0.9 & 0.9 & 1.8 & 0.9 & 0.9\\
$\kappa$ &20.0 & 3.0 & 3.0 & 3.3 & 3.0 & 3.0\\
$r_i$ & 4.5 & $\sim$ & $\sim$ & $\sim$ & $\sim$ & $\sim$\\
$r_{bw}$ & 15.70 & 5.05 & 4.61 & 4.04 & 5.02 & 5.49\\
$r_o$ & 41.36 & 41.44 & 41.55 & 15.71 & 15.62 & 15.53\\
$\gamma_s$ & 9.90 & 2.70 & 5.71 & 2.60 & 1.14 & 3.81\\
$\gamma_h$ & 12.28 & 3.10 & 6.34 & 2.31 & 0.78 & 3.55\\
$(\Delta \gamma)$ & 2.38 & 0.40 & 0.63 & -0.29 & -0.36 & -0.26\\
$N_{var}$ & $-$ & 1.88 & $-$ & $-$ & 0.30 & $-$\\
$\Delta d$ & $-$ & 8.80 & $-$ & $-$ & 5.26 & $-$\\
$\nu_{v,max} [Hz]$ & $-$ & 1.52 & $-$ & $-$ & 1.62 & $-$\\
$T_{d,max}$ [keV] & $-$ & 0.55 & $-$ & $-$ & 0.70 & $-$\\
$x_s [\%]$ & $-$ & 17 & $-$ & $-$ & 32 & $-$\\
$N_{shock}$ & $-$ & $-$ & 2.15 & $-$ & $-$ & 3.92\\
$\Delta r$ & $-$ & $-$ & 0.77 & $-$ & $-$ & 2.31\\
$D$ & 0.0 & 0.9 & 0.1 & 1.1 & 2.1 & 2.2\\
$x_{lag}$ & 7.8 & 3.1 & 4.2 & 3.0 & 2.0 & 2.7\\
$M [M_{\odot}]$ & 10.0 & $\sim$ & $\sim$ & $\sim$ & $\sim$ & $\sim$\\
$a_{*}$ & 0.5 & $\sim$ & $\sim$ & $\sim$ & $\sim$ & $\sim$\\
$n_H [10^{22} cm^{-2}]$ & 0.6 & $\sim$ & $\sim$ & $\sim$ & $\sim$ & $\sim$\\
$\chi^2_{red}$ & 2.73 & 2.49 & 2.90 & 6.74 & 5.55 & 5.57\\
$dof$ & 403.00 & 404.00 & 403.00 & 403.00 & 404.00 & 403.00\\
\end{tabular}
\end{table*}


\begin{thebibliography}{99}

\bibitem[\protect\citeauthoryear{Ar{\'e}valo 
\& Uttley}{2006}]{2006MNRAS.367..801A} Ar{\'e}valo P., Uttley P., 2006, MNRAS, 367, 801 

\bibitem[Arnaud(1996)]{1996ASPC..101...17A} Arnaud, K.~A.\ 1996, Astronomical Data Analysis Software and Systems V, 101, 17 

\bibitem[Axelsson et al.(2013)]{2013MNRAS.431.1987A} Axelsson, M., Hjalmarsdotter, L., \& Done, C.\ 2013, MNRAS, 431, 1987 

\bibitem[Belloni et 
al.(2005)]{2005A&A...440..207B} Belloni, T., Homan, J., Casella, P., et al.\ 2005, A\&A, 440, 207 

\bibitem[\protect\citeauthoryear{Belloni}{2010}]{2010LNP...794...53B} 
Belloni T.~M., 2010, LNP, 794, 53 

\bibitem[Campana \& Stella(1995)]{1995MNRAS.272..585C} Campana, S., \& Stella, L.\ 1995, MNRAS, 272, 585 

\bibitem[\protect\citeauthoryear{Casella}{2005}]{2005ApJ...629..403C}
Casella P., Belloni T., Stella L., 2005, ApJ, 629, 403 

\bibitem[Churazov et al.(2001)]{2001MNRAS.321..759C} Churazov, E., 
Gilfanov, M., \& Revnivtsev, M.\ 2001, MNRAS, 321, 759 

\bibitem[Cui et al.(1999)]{1999ApJ...512L..43C} Cui, W., Zhang, S.~N., Chen, W., \& Morgan, E.~H.\ 1999, ApJ, 512, L43 

\bibitem[\protect\citeauthoryear{Done, Gierli{\'n}ski, 
\& Kubota}{2007}]{2007A&ARv..15....1D} Done C., Gierli{\'n}ski M., Kubota A.,
2007, A\&ARv, 15, 1 

\bibitem[Ebisawa et al.(1994)]{1994PASJ...46..375E} Ebisawa, K., Ogawa, M., 
Aoki, T., et al.\ 1994, PASJ, 46, 375

\bibitem[Esin et al.(1997)]{1997ApJ...489..865E} Esin, A.~A., McClintock, 
J.~E., \& Narayan, R.\ 1997, ApJ, 489, 865 

\bibitem[Evans et 
al.(2007)]{2007A&A...469..379E} Evans, P.~A., Beardmore, A.~P., Page, K.~L., et al.\ 2007, A\&A, 469, 379 

\bibitem[Fragile et al.(2007)]{2007ApJ...668..417F} Fragile, P.~C., Blaes, 
O.~M., Anninos, P., \& Salmonson, J.~D.\ 2007, ApJ, 668, 417 

\bibitem[Fragile \& Blaes(2008)]{2008ApJ...687..757F} Fragile, P.~C., \& Blaes, O.~M.\ 2008, ApJ, 687, 757-766 

\bibitem[\protect\citeauthoryear{Frank, King, 
\& Raine}{2002}]{2002apa..book.....F} Frank J., King A., Raine D.~J.,
Accretion power in astrophysics, 3rd edition, 2002, Cambridge University Press 

\bibitem[Gehrels et al.(2004)]{2004ApJ...611.1005G} Gehrels, N., 
Chincarini, G., Giommi, P., et al.\ 2004, ApJ, 611, 1005 

\bibitem[Gierli{\'n}ski \& Done(2003)]{2003MNRAS.342.1083G} Gierli{\'n}ski, M., \& Done, C.\ 2003, MNRAS, 342, 1083 

\bibitem[Gilfanov et al.(2000)]{2000sgwa.work..114G} Gilfanov, M., Churazov, E., \& Revnivtsev, M.\ 2000, Proceedings of 5-th Sino-German workshop on Astrohpysics, 1999, Eds.~Gang Zhao, Jun-Jie Wang, Hong Mei Qiu and Gerhard Boerner, SGSC Conference Series, vol.1, pp.114-123, 114 

\bibitem[\protect\citeauthoryear{Gilfanov}{2010}]{2010LNP...794...17G} 
Gilfanov M., 2010, Lecture Notes in Physics, Vol. 794, The Jet Paradigm. Springer-Verlag, Berlin, p. 17 

\bibitem[Henisey et al.(2012)]{2012ApJ...761...18H} Henisey, K.~B., Blaes, O.~M., \& Fragile, P.~C.\ 2012, ApJ, 761, 18 

\bibitem[Homan et al.(2001)]{2001ApJS..132..377H} Homan, J., Wijnands, R., 
van der Klis, M., et al.\ 2001, ApJS, 132, 377 

\bibitem[Ingram et al.(2009)]{2009MNRAS.397L.101I} Ingram, A., Done, C., \& Fragile, P.~C.\ 2009, MNRAS, 397, L101 

\bibitem[\protect\citeauthoryear{Ingram 
\& Done}{2011}]{2011MNRAS.415.2323I} Ingram A., Done C., 2011 (ID11), MNRAS,
415, 2323 

\bibitem[\protect\citeauthoryear{Ingram 
\& Done}{2012}]{2012MNRAS.419.2369I} Ingram A., Done C., 2012 (ID12),
MNRAS, 419, 2369 

\bibitem[\protect\citeauthoryear{Ingram 
\& van der Klis}{2013}]{2013MNRAS.434.1476I} Ingram A., van der Klis M., 2013 (IK13),
MNRAS, 434, 1476 

\bibitem[Ingram(2016)]{2016AN....337..385I} Ingram, A.~R.\ 2016, Astronomische Nachrichten, 337, 385 

\bibitem[Ingram et al.(2016)]{2016MNRAS.461.1967I} Ingram, A., van der Klis, M., Middleton, M., et al.\ 2016, MNRAS, 461, 1967 

\bibitem[Jahoda et al.(1996)]{1996SPIE.2808...59J} Jahoda, K., Swank, 
J.~H., Giles, A.~B., et al.\ 1996, Proc. SPIE, 2808, 59 

\bibitem[Kotov et al.(2001)]{2001MNRAS.327..799K} Kotov, O., Churazov, E., \& Gilfanov, M.\ 2001, MNRAS, 327, 799 

\bibitem[Kubota \& Done(2004)]{2004MNRAS.353..980K} Kubota, A., \& Done, C.\ 2004, MNRAS, 353, 980 

\bibitem[Levine et al.(1996)]{1996AAS...189.3511L} Levine, A.~M., Cui, W., Remillard, R., et al.\ 1996, Bulletin of the American Astronomical Society, 28, 35.11 

\bibitem[Lyubarskii(1997)]{1997MNRAS.292..679L} Lyubarskii, Y.~E.\ 1997, 
MNRAS, 292, 679 

\bibitem[Mitsuda et al.(1984)]{1984PASJ...36..741M} Mitsuda, K., Inoue, H., 
Koyama, K., et al.\ 1984, PASJ, 36, 741 

\bibitem[Miyamoto et al.(1991)]{1991ApJ...383..784M} Miyamoto, S., Kimura, K., Kitamoto, S., Dotani, T., \& Ebisawa, K.\ 1991, ApJ, 383, 784 

\bibitem[Rapisarda et al.(2014)]{2014MNRAS.440.2882R} Rapisarda, S., 
Ingram, A., \& van der Klis, M.\ 2014 (RIK14), MNRAS, 440, 2882 

\bibitem[Rapisarda et al.(2016)]{2016MNRAS.462.4078R} Rapisarda, S., Ingram, A., Kalamkar, M., \& van der Klis, M.\ 2016 (RIKK16), MNRAS, 462, 4078 

\bibitem[Remillard et al.(2002)]{2002ApJ...564..962R} Remillard, R.~A., Sobczak, G.~J., Muno, M.~P., \& McClintock, J.~E.\ 2002, ApJ, 564, 962 

\bibitem[Remillard 
\& McClintock(2006)]{2006ARA&A..44...49R} Remillard, R.~A., \& McClintock, J.~E.\ 2006, ARA\&A, 44, 49 

\bibitem[Scaringi et al.(2012)]{2012MNRAS.427.3396S} Scaringi, S., K{\"o}rding, E., Uttley, P., et al.\ 2012, MNRAS, 427, 3396 

\bibitem[Scaringi(2014)]{2014MNRAS.438.1233S} Scaringi, S.\ 2014, MNRAS, 438, 1233 

\bibitem[Scaringi et al.(2015)]{2015SciA....1E0686S} Scaringi, S., Maccarone, T.~J., Kording, E., et al.\ 2015, Science Advances, 1, e1500686 

\bibitem[\protect\citeauthoryear{Shakura 
\& Sunyaev}{1973}]{1973A&A....24..337S} Shakura N.~I., Sunyaev R.~A., 1973,
A\&A, 24, 337 

\bibitem[Smith(1998)]{1998IAUC.7008....1S} Smith, D.~A.\ 1998, IAU Circ., 7008, 1 

\bibitem[Sobczak et al.(2000)]{2000ApJ...544..993S} Sobczak, G.~J., McClintock, J.~E., Remillard, R.~A., et al.\ 2000, ApJ, 544, 993 

\bibitem[Stella 
\& Vietri(1998)]{1998ApJ...492L..59S} Stella, L., \& Vietri, M.\ 1998, ApJ, 492, L59 

\bibitem[Sunyaev 
\& Truemper(1979)]{1979Natur.279..506S} Sunyaev, R.~A., \& Truemper, J.\ 1979, Nature, 279, 506 

\bibitem[Svensson 
\& Zdziarski(1994)]{1994ApJ...436..599S} Svensson, R., \& Zdziarski, A.~A.\ 1994, ApJ, 436, 599 

\bibitem[Thorne 
\& Price(1975)]{1975ApJ...195L.101T} Thorne, K.~S., \& Price, R.~H.\ 1975, ApJ, 195, L101 

\bibitem[Uttley 
\& McHardy(2001)]{2001MNRAS.323L..26U} Uttley, P., \& McHardy, I.~M.\ 2001, MNRAS, 323, L26 

\bibitem[Uttley et al.(2005)]{2005MNRAS.359..345U} Uttley, P., McHardy, 
I.~M., \& Vaughan, S.\ 2005, MNRAS, 359, 345 

\bibitem[van der Klis(2004)]{2004astro.ph.10551V} van der Klis, M.\ 2004, arXiv:astro-ph/0410551 

\bibitem[Walton et al.(2013)]{2013ApJ...777L..23W} Walton, D.~J., Zoghbi, A., Cackett, E.~M., et al.\ 2013, ApJ, 777, L23 

\bibitem[Wijnands 
\& van der Klis(1998)]{1998ApJ...507L..63W} Wijnands, R., \& van der Klis, M.\ 1998, ApJ, 507, L63 

\bibitem[Wilkinson 
\& Uttley(2009)]{2009MNRAS.397..666W} Wilkinson, T., \& Uttley, P.\ 2009, MNRAS, 397, 666 

\bibitem[Zdziarski et al.(1996)]{1996MNRAS.283..193Z} Zdziarski, A.~A., Johnson, W.~N., \& Magdziarz, P.\ 1996, MNRAS, 283, 193 

\bibitem[Zycki et al.(1999)]{1999MNRAS.305..231Z} Zycki, P.~T., Done, C., \& Smith, D.~A.\ 1999, MNRAS, 305, 231 


\end{thebibliography}
\end{document}